# Competitive ride-sourcing market with a third-party integrator

Yaqian Zhou[a], Hai Yang[a], Jintao Ke[b*], Hai Wang[c,d], Xinwei Li[e]

[a] Department of Civil and Environmental Engineering, The Hong Kong University of Science and Technology, Clear Water Bay, Kowloon, Hong Kong, China
[b] Department of Logistics and Maritime Studies, The Hong Kong Polytechnic University, Hung Hom, Kowloon, Hong Kong, China
[c] School of Information Systems, Singapore Management University, Singapore
[d] Heinz School of Information Systems and Public Policy, Carnegie Mellon University, Pittsburgh, United States
[e] School of economics and management, Beihang University, Beijing, China

**Abstract.**

Recently, some transportation service providers attempt to integrate the ride services offered by multiple independent ride-sourcing platforms, and passengers are able to request ride through such third-party integrators or connectors and receive service from any one of the platforms. This novel business model, termed as third-party *platform-integration* in this paper, has potentials to alleviate the cost of market fragmentation due to the demand splitting among multiple platforms. While most existing studies focus on the operation strategies for one single monopolist platform, much less is known about the competition and platform-integration as well as the implications on operation strategy and system efficiency. In this paper, we propose mathematical models to describe the ride-sourcing market with multiple competing platforms and compare system performance metrics between two market scenarios, i.e., with and without platform-integration, at Nash equilibrium as well as social optimum. We find that platform-integration can increase total realized demand and social welfare at both Nash equilibrium and social optimum, but may not necessarily generate a greater profit when vehicle supply is sufficiently large or/and market is too fragmented. We show that the market with platform-integration generally achieves greater social welfare. On one hand, the integrator in platform-integration is able to generate a thicker market and reduce matching frictions; on the other hand, multiple platforms are still competing by independently setting their prices, which help to mitigate monopoly mark-up as in the monopoly market.

---

[*] Corresponding author: Jintao Ke. (E-mail: jintao.ke@polyu.edu.hk).

Address: Room M624, 6/F, Li Ka Shing Tower, The Hong Kong Polytechnic University, Hung Hom, Kowloon, Hong Kong.

# 1. Introduction

Over the past few years, the rise of ride-sourcing services has disruptively reshaped the way people travel and made substantial impacts on traditional taxi industry and multimodal transportation system. Due to the low barrier under less restrictive entry permit, multiple ride-sourcing companies co-exist and compete with each other in many local markets. For example, Didi and Uber had been involved in fierce competition in China until the end of 2016 when Uber China was acquired by Didi. After that, although Didi accounts for the largest market share in China, it continuously faces challenges from new rivals, such as Meituan which expands its ride-sourcing services to a handful of cities in eastern China. By 2019, there are competitions between ride-sourcing companies all over the world, including Uber and Lyft in the United States, Grab and Go-Jek in Southeastern Asia, Uber and Bolt in Europe, Uber and Didi in Australia, Careem and Uber in the Middle East, Ola and Uber in India, and 99 Taxi, Cabify, and Uber in Brazil (Wang and Yang, 2019).

Competition is a double-edged sword for ride-sourcing markets. On one hand, similar to other service markets, competition in the ride-sourcing markets prevents a monopolist platform from greedily maximizing its own profit by distorting the operating strategies from the system efficient levels. On the other hand, competition between platforms leads to market fragmentation and increases matching frictions. Specifically, matching between demand and supply in ride-sourcing markets generally exhibits increasing returns to scale, namely, matching is more efficient as more passengers and drivers are involved in one platform, and therefore it is expected that the demand splitting between multiple independent platforms may result in certain market inefficiency.

Recently, there arises a novel business model, termed as *platform-integration*, which has potentials to address the market fragmentation under competition. It integrates ride-sourcing services offered by multiple companies in one APP interface as an integrator. For example, Baidu Map integrates the services of Didi and some smaller ride-sourcing companies into its APP and offer its users a wide variety of choices in China (Song, 2019). Without platform-integration, a passenger often requests a ride on one platform and thus can only be matched with an idle driver affiliated to that specific platform; with platform-integration, a passenger is able to request a ride through the platform integrator and can be matched with an idle driver affiliated to any of the integrated ride-sourcing platforms (e.g. Didi or Meituan). As a result, the accessible supply level for a passenger is higher on the integrator, and the average pick-up distance/time is expected to decline.

In actual operations, passengers may have freedom to choose the full set of ride-sourcing platforms displayed in the integrator (referred to as full platform-integration), or a subset of these platforms (referred to as partial platform-integration). In this paper, for analytical tractability, we restrict our analysis to the case of full platform-integration. It is worth noting that platform-integration is a novel business model that differs from *multi-homing* in that, passengers who opt for platform-integration can request ride on multiple platforms through one integrator interface, while passengers who opt for multi-homing patronize different platforms by switching APPs (once a driver is dispatched from a specific platform, passengers need to cancel requests on other platforms). On the supply side, drivers in market scenario with platform-integration are still affiliated to only one specific platform and get paid based on its (i.e., that specific platform's) corresponding pricing and incentives strategy, while drivers who opt for multi-homing are simultaneously affiliated to more than one platforms and get paid from the platform from which an order is dispatched. It is also worth noting that the platform-integration is different from simply merging the platforms in competition. In market scenario with platform-integration, the participating platforms still have freedom to determine their trip fares and other operating strategies while sharing some information (e.g. locations of idle drivers) to the integrator. By contrast, in the scenario of platform merging, some platforms are turned into one single platform to get competitive advantage and no longer make individual pricing and operating decisions; in an extreme case where all platforms are merged, the market essentially becomes a monopoly market. In a sense, the platform-integration is able to (1) maintain the competition between platforms and prevent them greedily grapping too much profits and (2) reduce the market fragmentation in the meantime.

Of particular interest is to understand the implications of platform-integration on platform's operating strategies and the resulting passenger demand, platform profit, and social welfare. Platform-integration is generally expected to improve system efficiency with increased market thickness on the integrator. However, the competing and integrated platforms may adjust their operating strategies (e.g., trip fare) in response to the potential change of system efficiency, which may in turn affect the market equilibrium and the degree of efficiency improvement. This intricate feedback loops raise the need for trackable mathematical models.

To address these challenges, in this paper we propose mathematical models to describe the ride-sourcing market with multiple competing platforms, and compare vehicle utilization rate, realized demand, the profit and social welfare in two markets scenarios: (1) competitive market without platform-integration; and (2) competitive market with platform-integration. We prove that platform-integration is able to increase total realized demand and social welfare at both Nash equilibrium and social optimum, since it can eliminate the market fragmentation cost but still maintain the competitions to avoid the platforms arrogating profits from the market. However, platform-integration may not increase platform profit when the market is too fragmented (i.e., when the number of platforms is large), in which platforms may face severe competitions and are forced to set an extremely low trip fare. In addition, we conduct some numerical studies to investigate the effects of the commission fee charged by the integrator from the ride-sourcing platforms on the market measures in a general mixed market where some passengers use ride-sourcing services via the integrator, and others directly request orders through the ride-sourcing platforms. Useful managerial insights obtained from both theoretical and numerical studies can help the platforms and the government to design appropriate operating strategies for maximizing platform profit and social welfare, respectively.

The rest of the paper is organized as follows. Section 2 reviews the literature related to our study. Section 3 presents stationary equilibrium models and optimal operating strategies at Nash equilibrium or social optimum, in scenarios with and without platform-integration. Section 4 analytically compares the optimal strategies and the corresponding profit, consumer surplus and social welfare. Section 5 provides numerical studies to illustrate the theoretical results and demonstrate to what extent the supply and market fragmentation affect the performance of platform-integration, and investigate the impacts of commission fee in the general mixed market equilibrium; Section 6 concludes the paper and outlooks on future research.

## 2. Literature review

As pioneers in sharing economy, ride-sourcing services have attracted attentions from researchers in various disciplines, including transportation engineering, operations research, computer science, economics, and management science. General research issues investigated so far include market equilibrium analysis in both aggregate (Zha et al., 2016) and disaggregate settings (He et al., 2018); supply and demand coordination with price and wage (Bai et al., 2019; Taylor, 2018; Yang et al., 2020b); homogenous and/or heterogeneous price and wage over space and time (Cachon et al., 2015; Castillo et al., 2017; Bimpikis et al., 2019); matching and dispatching strategy (Xu et al., 2017; Zha et al., 2018; Xu et al., 2018; Lyu et al. 2019; Yang et al., 2020a); driver supply and elasticity with respect to income level and other factors (Chen and Sheldon, 2016; Zha et al., 2017; Sun et al., 2019a; Sun et al., 2019b); implications of autonomy and electrification on shared mobility services (Ke et al., 2019; Al-Kanj et al., 2020); integration of shared mobility services with public transit system and parking infrastructure (Xu et al., 2017; Zhu et al, 2020); ridepooling ride-sourcing services (Ke et al, 2020). Readers may refer to Wang and Yang (2019) for a comprehensive review of research problems in ride-sourcing markets. It is noteworthy that most of these studies focus on a market with one single monopolist platform that has the market power to achieve a monopoly optimum by tuning the trip fares, wages, and matching strategies.

So far, only a few studies have been devoted to the analysis of the ride-sourcing market with platform competitions. For example, Zha et al. (2016) study a duopoly market, where a Nash equilibrium is reached at which the two platforms cannot unilaterally change their decisions (trip fare and vehicle fleet size) to further increase their profits. They argue that competition does not necessarily lower the trip fare or increase the social welfare because competition segments the groups of passengers and drivers and thus increases matching frictions. Cohen and Zhang (2017) investigate a duopoly market in which two ride-sourcing platforms choose their trip fares charged to passengers and wages paid to drivers and compete for both demand and supply. Séjourné et al. (2019) show that the splitting of demand between different platforms in essence makes the market thinner, and thus leads to inefficiency resulting from market fragmentation.

There is a rich body of literature in economics on the general two-sided market (Rochet and Tirole, 2003; Armstrong, 2006; Armstrong and Wright, 2007) relevant to platform-integration. Some seminal models are extended to analyze the ride-sharing and ride-sourcing markets. For example, Jeitschko and Tremblay (2019) study the two-sided market at which passengers endogenously determine whether they only patronize one platform or simultaneously two competing platforms, and show that passengers prefer the platform which offers a lower price. Bernstein et al. (2019) examine the multi-homing on the supply side. Specifically, they analyze and compare two settings, one with drivers serving passengers on one platform, and the other with drivers serving passengers on both platforms. They find that, an individual driver may increase his/her income by choosing multi-homing, but no drivers will become better off if all of them choose multi-homing. In general, these studies develop stylized models to describe two groups of agents (i.e., demand and supply in markets) and offer interesting managerial insights. They do not characterize certain specific process in the ride-sourcing market, e.g., the matching between drivers and passengers, and thus may result in some biased conclusions in the ride-sourcing context.

## 3. Market equilibrium and optimal strategy

In this section, we first present a model to describe the equilibrium of a market without platform-integration, and then extend the model to delineate the equilibrium of a market with platform-integration. We study the optimal strategies for maximizing platform profit and social welfare respectively, namely, the Nash equilibrium and social optimum.

### 3.1. Market without platform-integration

Consider a ride-sourcing market with $I$ competing platforms ($i = 1, 2, \cdots, I$), a group of passengers and a group of drivers affiliated to only one specific platform. In the market without platform-integration, a driver only works for one particular platform and a passenger only sends ride request to one platform for a particular trip. Let $Q = \bar{Q}f(C)$ represent the demand function, where $\bar{Q}$ is maximum possible demand (i.e., the potential arrival rate of passengers), $f(\cdot)$ is a decreasing function of a generalized cost $C$, and $Q$ is realized passenger demand, i.e., the actual arrival rate of passengers in the ride-sourcing market regardless of which platform they choose. Then the inverse demand function (denoted by $B(Q)$) can be given by $B(Q) = f^{-1}(Q/\bar{Q})$. Let $q_i \geq 0$ denote the realized demand for platform $i$, then we have $Q = \sum_i q_i$. Passengers are assumed to be homogenous in terms of their value of time ($\beta$). Each platform $i$ chooses the trip fare $F_i$ to maximize its profit while advanced pricing features such as surge pricing are not considered. Let $W(N_i^v)$ denote the waiting and pick-up time experienced by passengers opting for platform $i$, which depends on the number of idle vehicles $N_i^v$ on the platform. Following Castillo et al. (2018), we assume that each platform implements a First-Come-First-Serve matching mechanism, and dispatches idle vehicle to a passenger immediately after he/she raises the travel request. In this case, the waiting time for matching is negligible, and $W(N_i^v)$ is dominated by the pick-up time. Let $N_i$ and $N_i^v$ denote the fleet size and number of idle vehicles in equilibrium on platform $i$, respectively; let $N$ and $N^v$, respectively, denote the total fleet size and total number of idle vehicles in equilibrium in the market, i.e., $N = \sum_i N_i$ and $N^v = \sum_i N_i^v$. Let $T$ denote the average trip time and is assumed to be equal among all platforms. A user equilibrium is reached when passengers are indifferent with respect to all platforms, that is, the generalized costs (constituted of the monetary cost and time cost) in all ride-sourcing platforms are equal. In equilibrium, for each platform, the vehicle fleet size equals the sum of the numbers of vacant vehicles ($N_i^v + q_i W(N_i^v)$) and occupied vehicles ($q_i T$).[2] Thus, we have the following equations:

$$B(Q) = F_i + \beta(T + W(N_i^v)), \forall i \tag{1}$$

$$N_i = N_i^v + q_i(T + W(N_i^v)), \quad \forall i \tag{2}$$

$$Q = \sum_{i=1}^{I} q_i \tag{3}$$

$$N^v = \sum_{i=1}^{I} N_i^v \tag{4}$$

$$N = \sum_{i=1}^{I} N_i \tag{5}$$

---

[2] Vacant vehicles consist of idle vehicles that are available and those on the way to pick up the assigned passengers.

Let $c$ denote the average operating cost of a vehicle per time unit. In the market without platform-integration, each platform decides its trip fare $F_i$. The optimal profit of each platform at Nash Equilibrium (NE), denoted as $P_1^{ne}, P_2^{ne}, \cdots, P_I^{ne}$, can be obtained using a set of nonnegative $q_i^{ne}$ that solves the following optimization problem for each platform $i$ (hereinafter, variables with a superscript "$ne$" denote the optimal solutions at Nash equilibrium).

$$\max_{q_i \geq 0} P_i = q_i F_i - c N_i = q_i \left( B(Q) - \beta(T + W(N_i^v)) \right) - c N_i \qquad (6)$$

s.t. Eqs(1) − (3)

Note that after substituting $F_i$ into the objective function by using the inverse demand function (1), searching the optimal passenger demand of each platform $q_i$ is equivalent to searching the optimal fare $F_i$. The solutions $(q_1^{ne}, q_2^{ne}, \cdots, q_I^{ne})$ can be obtained using the first-order conditions:

$$\left( B(Q^{ne}) + q_i^{ne} B'(Q^{ne}) + \beta \frac{dN_i^v}{dq_i} \Big|_{q_i = q_i^{ne}} \right) q_i^{ne} = 0, \qquad \forall i \qquad (7)$$

$$B(Q^{ne}) + q_i^{ne} B'(Q^{ne}) + \beta \frac{dN_i^v}{dq_i} \Big|_{q_i = q_i^{ne}} \leq 0, \qquad \forall i \qquad (8)$$

where

$$Q^{ne} = \sum_{i=1}^{I} q_i^{ne}, q_i^{ne} \geq 0 \qquad (9)$$

The social optimum scheme represents an ideal case where a social planner has full control of all the ride-sourcing platforms and aims to maximize social welfare (denoted as $S$) by deciding the trip fares. Similarly, after substituting $F_i$ into the objective function below by using the inverse demand function in Eq(1), the unconstrained welfare-maximizing problem can be formulated as:[3]

$$\max_{q_i \geq 0} S = \int_0^Q B(x) dx - \sum_{i=1}^{I} q_i B(Q) + \sum_{i=1}^{I} q_i F_i - cN = \int_0^Q B(x) dx - \sum_{i=1}^{I} q_i \beta(T + W(N_i^v)) - cN \qquad (10)$$

---

[3] Since the profits for ride-sourcing platforms may be in deficit at unconstrained welfare-maximizing scheme, one may constrain that the profits are nonnegative or at least equal to a reservation value, which makes the social optimum sustainable. In this paper, we focus on the social optimum without profit constraint.

set Eqs(1) − (3)

Let $(q_1^{so}, q_2^{so}, \cdots, q_I^{so})$ denote the optimal solution to (10) (hereinafter, superscript "$so$" stands for social optimum) which can be obtained using the following first-order conditions:

$$\left(B(Q^{so}) + \beta \frac{dN_i^v}{dq_i}\bigg|_{q_i=q_i^{so}}\right) q_i^{so} = 0, \qquad \forall i \tag{11}$$

$$B(Q^{so}) + \beta \frac{dN_i^v}{dq_i}\bigg|_{q_i=q_i^{so}} \leq 0, \qquad \forall i \tag{12}$$

where

$$Q^{so} = \sum_{i=1}^{I} q_i^{so}, q_i^{so} \geq 0 \tag{13}$$

Taking the first-order and second-order derivatives of both sides of Eq(2) with respect to $q_i$ yields:

$$\frac{dN_i^v}{dq_i} = -\frac{T + W_i}{1 + q_i W_i'}, \qquad \forall i \tag{14}$$

$$\frac{d^2 N_i^v}{dq_i^2} = \frac{(T + W_i)\left(2W_i' + q_i W_i'' \frac{dN_i^v}{dq_i}\right)}{(1 + q_i W_i')^2}, \qquad \forall i \tag{15}$$

where $W_i$, $W_i'$, respectively, represent $W(N_i^v)$ and $dW(N_i^v)/dN_i^v$ for simplicity.

Clearly, the signs of $dN_i^v/dq_i$ and $d^2 N_i^v/dq_i^2$ are uncertain. If $1 + q_i W_i' < 0$, the number of idle vehicles increases with demand, which is an inefficient market outcome termed as *Wild Goose Chases* (WGCs) (Castillo et al., 2018). In such case, the density of idle drivers is extremely low, which forces the platform to match passengers with distant drivers. Then drivers spend substantial time on the way to pick up passengers. It can be easily found from objective function (6) and the first-order condition (12) that the optimal solutions at both Nash equilibrium and social optimum in the market without platform-integration do not locate in the WGC regions.[4]

## 3.2. Market with platform-integration

Next we consider an alternative scenario with an integrator that enables passengers to access all $I$ competing platforms simultaneously. Hereinafter, we use "~" to denote the counterparts of variables with platform-integration. Realized passenger demand $\tilde{Q}$ are then divided into two groups: passengers who opt for the integrator (with an arrival rate of $\tilde{Q}_1$) and those who do not (with an arrival rate of $\tilde{Q}_2$), i.e., $\tilde{Q} = \tilde{Q}_1 + \tilde{Q}_2$. Accordingly, for each platform, its realized demand $\tilde{q}_i$ consist of those from the integrator (denoted by $\tilde{q}_{i_1}$) and those directly from its own platform (denoted as $\tilde{q}_{i_2}$). Then $\tilde{Q}_1 = \sum_{i=1}^{I} \tilde{q}_{i_1}$ and $\tilde{Q}_2 = \sum_{i=1}^{I} \tilde{q}_{i_2}$. Requests of passengers who order services via the integrator are assumed to be assigned to the ride-sourcing platforms depending on the availability of idle vehicles and corresponding pick-up distances without discrimination between platforms. In other words, when a passenger raises a request through the integrator, the integrator assigns the passenger to the nearest idle driver regardless of which platform the driver is affiliated to. Let $\widetilde{N}^v$ denote the total number of idle vehicles, and $\widetilde{N}_i^v$ the number of idle vehicles on platform $i$, then it should meet $\widetilde{N}^v = \sum_i \widetilde{N}_i^v$. The waiting and pickup time of passenger opting for the integrator is $W(\widetilde{N}^v)$, while that of platform $i$ is $W(\widetilde{N}_i^v)$. If platforms are free to re-optimize their trip fares after platform-integration is introduced, then a new Nash equilibrium may occur. Let $\tilde{F}_i$ and $\widetilde{\omega}_i$ denote the trip fare and market share of platform $i$ in the integrator (i.e., $\widetilde{\omega}_i = \tilde{q}_{i_1}/\tilde{Q}_1$), respectively, then the expected trip fare in the integrator is $\sum_{i=1}^{I} \widetilde{\omega}_i \tilde{F}_i$; Let $\tilde{\tau}$ denote the commission fee charged to the passengers by the integrator[5], while $\tilde{\tau} \geq 0$ indicates passengers pay fee to the integrator and $\tilde{\tau} < 0$ indicates the integrator offers some compensation to attract passengers. Then the monetary cost of passengers who opt for the integrator can be approximated by $\sum_{i=1}^{I} \widetilde{\omega}_i \tilde{F}_i + \tilde{\tau}$. Let $\tilde{T}$ denote the average trip time in the market with platform-integration and is assumed to be equal among all platforms. Then the generalized trip costs of passengers who opt for the integrator, denoted as $\tilde{C}_1$ and passengers who opt for platform $i$, denoted as $\tilde{C}_2$ in equilibrium, are given by, respectively:

$$\tilde{C}_1 = \sum_{i=1}^{I} \widetilde{\omega}_i \tilde{F}_i + \beta\left(\tilde{T} + W(\tilde{N}^v)\right) + \tilde{\tau} \tag{16}$$

and

$$\tilde{C}_2 = \tilde{F}_i + \beta\left(\tilde{T} + W(\tilde{N}_i^v)\right) \tag{17}$$

---

[4] Clearly, it should meet $\beta dN_i^v/dq_i|_{q_i=q_i^{wm}} \leq 0$ due to the fact that $B(Q^{wm}) \geq 0$, which implies that social optimal solutions do not locate in the WGC regions. Given the optimal passenger demand $q_i^{pm}$, it always satisfies the number of idle vehicles in the normal region is larger than its counterpart in the WGC region due to the inverted-U shape of $f_i: N_i^v \to q_i$ (see Figure 6 in **Appendix A** for reference). Then, considering the objective function (6) at Nash equilibrium, the lower waiting time (or equivalently, the higher trip fare) in the normal region leads to a greater profit in the normal region than in the WGC region. Thus, the optimal solutions at Nash equilibrium are not located in the WGC region either.

[5] At present, the platform integrators such as Baidu and Gaode do not impose commission charge neither on passengers nor on individual platforms (but benefit indirectly from passengers' patronage of their free services). Here, for generality, we suppose a commission charge (it can be positive, zero, or negative) is imposed on passengers rather the platform.

Suppose every passenger chooses the mode that minimizes its trip cost at equilibrium. Without loss of generality, let $k$ denote the index for the platform charging the highest trip fare, i.e., $\tilde{F}_k = \max\{\tilde{F}_i\}, \forall i$; let $j$ denote the index for the platform charging the lowest trip fare, i.e., $\tilde{F}_j = \min\{\tilde{F}_i\}, \forall i$. Then, if $\tilde{\tau} < \tilde{F}_j - \tilde{F}_k$, then $\tilde{C}_1 - \tilde{C}_2 < \sum_{i=1}^{I} \widetilde{\omega}_i \tilde{F}_i + \tilde{\tau} - \tilde{F}_j < \tilde{F}_k + \tilde{F}_j - \tilde{F}_k - \tilde{F}_j = 0$, which indicates the generalized trip cost of the passengers who opt for the integrator is always lower than that of passengers who do not, hence all passengers will choose the integrator. If $\tilde{\tau} > \beta W(\tilde{N}_j^v)$, then $\tilde{C}_1 - \tilde{C}_2 > \sum_{i=1}^{I} \widetilde{\omega}_i \tilde{F}_i + \beta \tilde{T} + \tilde{\tau} - \tilde{C}_2 > \tilde{F}_j + \beta\left(\tilde{T} + W(\tilde{N}_j^v)\right) - \tilde{C}_2 = 0$, which indicates the generalized trip cost of passengers who opt for the integrator is always higher than that of the passengers who do not, hence no passengers choose the integrator. Intuitively, if $\tilde{\tau}$ is sufficiently small (e.g., below a threshold $\tilde{\tau}_1$) such that $\tilde{C}_1 < \tilde{C}_2$ always holds, then all passengers choose the integrator; if $\tilde{\tau}$ is sufficiently large (e.g., above a threshold $\tilde{\tau}_2$) such that $\tilde{C}_1 > \tilde{C}_2$ always holds, then no passengers choose the integrator; if $\tilde{\tau}$ is medium (e.g., between $\tilde{\tau}_1$ and $\tilde{\tau}_2$) with $\tilde{C}_1 = \tilde{C}_2$, passengers may be indifferent between the $I$ independent platforms and the integrator. Note that $\tilde{\tau}_1$ and $\tilde{\tau}_2$ depend on many factors, including exogenous variables (e.g., vehicle fleet size), platform decisions (i.e., trip fare), inverse demand function $B(\cdot)$ and waiting time $W(\cdot)$. It is possible that there exist no non-negative $\tau_1$ such that all passengers choose the integrator when vehicle supply is sufficiently large or/and vehicle fleet sizes vary greatly among platforms. Based on discussion above, we then have the following equations:

$$B(\tilde{Q}) = \min(\tilde{C}_1, \tilde{C}_2) \tag{18}$$

$$\tilde{q}_{i_1}(\tilde{C}_2 - \tilde{C}_1) \geq 0, \quad \forall i \tag{19}$$

$$\tilde{q}_{i_2}(\tilde{C}_1 - \tilde{C}_2) \geq 0, \quad \forall i \tag{20}$$

where constraints (19) and (20), respectively, indciate $\tilde{q}_{i_1} = 0$ if $\tilde{C}_1 > \tilde{C}_2$ (i.e., no passengers choose the integrator) and $\tilde{q}_{i_2} = 0$ if $\tilde{C}_2 > \tilde{C}_1$ (i.e., all passengers choose the integrator). In equilibrium, the number of rides $(\tilde{q}_{i_1}, \tilde{q}_{i_2})$ that platform $i$ serves should also meet the following vehicle conservation constraint:

$$N_i = \tilde{N}_i^v + \tilde{q}_{i_1}\left(\tilde{T} + W(\tilde{N}^v)\right) + \tilde{q}_{i_2}\left(\tilde{T} + W(\tilde{N}_i^v)\right), \quad \forall i \tag{21}$$

Under platform-integration, the supply (i.e., idle vehicles) of the competing platforms are managed in one matching pool without discrimination and the integrator assigns the passenger to the nearest idle driver. In that case, for each platform, its realized demand $\tilde{q}_{i_1}$ for the integrator should be proportional to its number of idle $\tilde{N}_i^v$, or equivalently, it should meet:

$$\frac{\tilde{q}_{i_1}}{\tilde{q}_{j_1}} = \frac{\tilde{N}_i^v}{\tilde{N}_j^v}, \quad \forall i,j \tag{22}$$

For further analysis, we make the following two mild assumptions.

**Assumption 1.** *The average trip times on each platform with and without a platform integrator are equal, i.e., $T = \tilde{T}$.*

This assumption is realistic since vehicles on different platforms are operating on the same road network under the same traffic condition.

**Assumption 2.**

*(a) The waiting and pick-up time function $W(\cdot)$ is convex, strictly decreasing with the number of idle vehicles and continuously differentiable.*

*(b) The inverse demand function $B(\cdot)$ is convex, strictly decreasing with the realized passenger demand and continuously differentiable, and $Q \cdot B(Q)$ is concave for $Q > 0$.*

*(c) The total realized demand at both Nash equilibrium and social optimum is positive, i.e., $Q^{ne} > 0, Q^{so} > 0, \tilde{Q}^{ne} > 0, \tilde{Q}^{so} > 0$.*

**Assumption 2-(a)** says that the waiting and pick-up time monotonically decreases with the number of idle vehicles, but the declining slope (representing the marginal decrease caused by a unit increase of idle vehicles) becomes flatter as the number of idle vehicles increases. **Assumption 2-(b)** is required for the existence and uniqueness of the optimization solutions at either Nash equilibrium or social optimum, and is usually adopted for demand function in revenue management. **Assumption 2-(c)** is to prevent unrealistic cases with $Q^{pm} = 0, Q^{wm} = 0, \tilde{Q}^{pm} = 0, \tilde{Q}^{wm} = 0$.

If commission fee $\tilde{\tau}$ charged by the integrator is lower than $\tilde{\tau}_1$, passengers pay equal expected trip fare (given by $\sum_{i=1}^{I} \widetilde{\omega}_i \tilde{F}_i$) and bear equal waiting and pick-up time (given by $W(\tilde{N}^v)$) because all of them raise requests through the integrator. Then Eqs(18)-(22) reduce to:

$$B(\tilde{Q}) = \sum_{i=1}^{I} \widetilde{\omega}_i \tilde{F}_i + \beta\left(\tilde{T} + W(\tilde{N}^v)\right) + \tilde{\tau} \tag{23}$$

$$N_i = \tilde{N}_i^v + \tilde{q}_i\left(\tilde{T} + W(\tilde{N}^v)\right), \quad \forall i \tag{24}$$

$$\tilde{N}^v = \sum_{i=1}^{I} \tilde{N}_i^v \tag{25}$$

$$\tilde{Q} = \sum_{i=1}^{I} \tilde{q}_i \tag{26}$$

$$\frac{\tilde{q}_i}{\tilde{q}_j} = \frac{\tilde{N}_i^v}{\tilde{N}_j^v}, \quad \forall i,j \tag{27}$$

Using vehicle conservation constraint (24) as well as demand split rule for the integrator (27), we further obtain $(\tilde{N}_i^v + \tilde{q}_i(\tilde{T} + W(\tilde{N}^v)))/\tilde{N}_i^v = (\tilde{N}_j^v + \tilde{q}_j(\tilde{T} + W(\tilde{N}^v)))/\tilde{N}_j^v$, then $N_i/\tilde{N}_i^v = N_j/\tilde{N}_j^v$, that is, the number of idle vehicles on each platform is proportional to its corresponding vehicle fleet size, leading to the following relationship:

$$\tilde{N}_i^v = \frac{N_i}{N}\tilde{N}^v, \quad \forall i \tag{28}$$

If $\tilde{\tau}$ lies between $\tilde{\tau}_1$ and $\tilde{\tau}_2$, passengers who opt for the integrator and those who opt for platform $i$ should have the same generalized trip costs in equilibrium. Then Eqs(16)-(18) are equivalent to the following equation:

$$B(\tilde{Q}) = \sum_{i=1}^{I} \widetilde{\omega}_i \tilde{F}_i + \beta\left(\tilde{T} + W(\tilde{N}^v)\right) + \tilde{\tau} = \tilde{F}_i + \beta\left(\tilde{T} + W(\tilde{N}_i^v)\right) \tag{29}$$

If $\tilde{\tau}$ is above $\tilde{\tau}_2$, no passengers choose the integrator due to its high commission charge, then the market is exactly equivalent to that without platform-integration, which has been analyzed in the previous section.

In the market with platform-integration, each platform $i$ chooses $\tilde{F}_i$ to maximize its own profit (denoted as $\tilde{P}_i$) at Nash equilibrium.[6] Note that in the case of $\tilde{\tau} < \tilde{\tau}_1$, all trip fare sets $(\tilde{F}_1, \tilde{F}_2, \cdots, \tilde{F}_I)$ that do not satisfy $\tilde{F}_i = \tilde{F}_j = \sum_{i=1}^{I} \widetilde{\omega}_i \tilde{F}_i, \forall i, j$ are not solutions at equilibrium, because the platform charging lower price has incentive to increase its price to reap greater profit. In other words, all platforms must set the equal trip fare (denoted as $\tilde{F}$) to achieve a stable market equilibrium. After substituting $\tilde{F}_i$ into the objective function below by using the inverse demand function (23), together with $\tilde{F}_i = \tilde{F}_j = \sum_{i=1}^{I} \widetilde{\omega}_i \tilde{F}_i, \forall i, j$, the Nash equilibrium (NE) in the case of $\tilde{\tau} < \tilde{\tau}_1$ with platform-integration can be formulated as:[7]

$$\max_{\tilde{q}_i \geq 0} \tilde{P}_i = \tilde{q}_i \tilde{F}_i - cN_i = \tilde{q}_i \left( B(\tilde{Q}) - \beta \left( \tilde{T} + W(\tilde{N}^v) \right) - \tilde{\tau} \right) - cN_i \tag{30}$$

s.t. Eqs(23)-(28)

Let $(\tilde{q}_1^{ne}, \tilde{q}_2^{ne}, \cdots, \tilde{q}_I^{ne})$ be a Nash equilibrium solution in the market with platform-integration, then the following first-order conditions must be satisfied:

$$\left( B(\tilde{Q}^{ne}) + \tilde{q}_i^{pm} B'(\tilde{Q}^{ne}) + \frac{N_i}{N} \beta \frac{d\tilde{N}^v}{d\tilde{q}_i} \Big|_{\tilde{q}_i = \tilde{q}_i^{ne}} - \tilde{\tau} \right) \tilde{q}_i^{ne} = 0, \quad \forall i \tag{31}$$

$$B(\tilde{Q}^{ne}) + \tilde{q}_i^{ne} B'(\tilde{Q}^{ne}) + \frac{N_i}{N} \beta \frac{d\tilde{N}^v}{d\tilde{q}_i} \Big|_{\tilde{q}_i = \tilde{q}_i^{ne}} - \tilde{\tau} \leq 0, \quad \forall i \tag{32}$$

where

---

[6] Here for generality, we suppose each individual platform is able to re-optimizes its fare after joining the integrator. In practice, platforms may keep their pricing strategies unchanged with the integrator, so we also compare the case with unchanged fare in the presence of platform-integration through numerical studies.

[7] It is worth mentioning that the profit-maximizing problem in the monopoly market can be mathematically formulated as $\max_{Q>0} P = QF - cN = Q\left( B(Q) - \beta(T + W(N^v)) \right) - cN$, which is exactly the same as maximizing sum of the profit of each platform in the market with platform-integration, defined in (30). Therefore, the solutions at monopoly optimum may be inconsistent with those obtained in the market with platform-integration at Nash equilibrium, $I \geq 2$.

$$\tilde{Q}^{ne} = \sum_{i=1}^{I} \tilde{q}_i^{ne}, \tilde{q}_i^{ne} \geq 0 \tag{33}$$

Social welfare in the market with platform-integration (denoted as $\tilde{S}$) is defined to be the sum of consumer surplus and profits of all platforms and the integrator. In the case of $\tilde{\tau} < \tilde{\tau}_1$, the objective function at social optimum can be mathematically written as:

$$\max_{\tilde{q}_i \geq 0} \tilde{S} = \int_0^{\tilde{Q}} B(x)dx - \sum_{i=1}^{I} \tilde{q}_i \beta \left(\tilde{T} + W(\tilde{N}^v)\right) - cN \tag{34}$$

s.t. Eqs(23)-(28)

Let $(\tilde{q}_1^{so}, \tilde{q}_2^{so}, \cdots, \tilde{q}_I^{so})$ denote the optimal solution to (34), then the following first-order conditions hold:

$$\left(B(\tilde{Q}^{so}) + \frac{N_i}{N}\beta \frac{d\tilde{N}^v}{d\tilde{q}_i}|_{\tilde{q}_i=\tilde{q}_i^{so}}\right)\tilde{q}_i^{so} = 0, \quad \forall i \tag{35}$$

$$B(\tilde{Q}^{so}) + \frac{N_i}{N}\beta \frac{d\tilde{N}^v}{d\tilde{q}_i}|_{\tilde{q}_i=\tilde{q}_i^{so}} \leq 0, \quad \forall i \tag{36}$$

where

$$\tilde{Q}^{so} = \sum_{i=1}^{I} \tilde{q}_i^{so}, \tilde{q}_i^{so} \geq 0 \tag{37}$$

From Eq(23), we can see that the optimal solution $(\tilde{q}_1^{so}, \tilde{q}_2^{so}, \cdots, \tilde{q}_I^{so})$ is independent of the commission fee charged by the integrator, because trip fare and commission fee can be packaged as an auxiliary decision variable for the social planner. Taking the first-order and second-order derivatives of both sides of Eq(24) with respect to $\tilde{q}_i$ after substituting Eq(28) into Eq(24) yields:

$$\frac{d\tilde{N}^v}{d\tilde{q}_i} = -\frac{\tilde{T} + \tilde{W}}{\frac{N_i}{N} + \tilde{q}_i \tilde{W}'}, \quad \forall i \tag{38}$$

$$\frac{d^2\tilde{N}^v}{d\tilde{q}_i^2} = \frac{(\tilde{T} + \tilde{W})\left(2\tilde{W}' + \tilde{q}_i \tilde{W}'' \frac{d\tilde{N}^v}{d\tilde{q}_i}\right)}{\left(\frac{N_i}{N} + \tilde{q}_i \tilde{W}'\right)^2}, \quad \forall i \tag{39}$$

where $\tilde{W}$, $\tilde{W}'$, respectively, represent $W(\tilde{N}^v)$ and $dW(\tilde{N}^v)/d\tilde{N}^v$ for simplicity.

The sign of $d\widetilde{N}^v/d\tilde{q}_i$ and $d^2\widetilde{N}^v/d\tilde{q}_i^2$ are undetermined because the sign of the term $N_i/N + \tilde{q}_i\widetilde{W}'$ is indeterminate. If $N_i/N + \tilde{q}_i\widetilde{W}' > 0$, the number of idle vehicles decreases with the passenger demand, which indicates the normal non-WGC regime. From objective function (30) and first-order condition (36), we can prove that the optimal solutions at both Nash equilibrium and social optimum in the market with platform-integration are not in the WGC region.

By summing up both sides of $I$ equations in Eq(24), we also obtain:

$$N = \widetilde{N}^v + \widetilde{Q}(\widetilde{T} + \widetilde{W}) \tag{40}$$

We then take the first order and second-order derivatives of both sides of Eq(40) with respect to $\widetilde{Q}$, which gives rise to:

$$\frac{d\widetilde{N}^v}{d\widetilde{Q}} = -\frac{\widetilde{T} + \widetilde{W}}{1 + \widetilde{Q}\widetilde{W}'} \tag{41}$$

$$\frac{d^2\widetilde{N}^v}{d\widetilde{Q}^2} = \frac{(\widetilde{T} + \widetilde{W})\left(2\widetilde{W}' + \widetilde{Q}\widetilde{W}''\frac{d\widetilde{N}^v}{d\widetilde{Q}}\right)}{(1 + \widetilde{Q}\widetilde{W}')^2} \tag{42}$$

It is worth noting that problem (34) can be reformulated with regard to $\widetilde{Q}$ as follows:

$$\max_{\widetilde{Q}>0} S = \int_0^{\widetilde{Q}} B(x)dx - \widetilde{Q}\beta(\widetilde{T} + \widetilde{W}) - cN \tag{43}$$

From constraint (40), we can see the maximum social welfare in the market with platform-integration is also independent of the number of platforms $I$.[8] Noted that $\widetilde{Q}^{wm} > 0$ by **Assumption 2-(c)**, then the first-order condition for problem (43) with an interior solution should satisfy:

$$B(\widetilde{Q}^{so}) + \beta\frac{d\widetilde{N}^v}{d\widetilde{Q}}|_{\widetilde{Q}=\widetilde{Q}^{so}} = 0 \tag{44}$$

---

[8] Based on this, when the supply capacity $N$ (i.e., vehicle fleet sizes) is fixed, the solutions in the monopoly market at social optimum exactly coincides with social optimal solutions in the market with platform-integration in the case of $\tilde{\tau} < \tilde{\tau}_1$.

In the case that $\tilde{\tau}$ lies between $\tilde{\tau}_1$ and $\tilde{\tau}_2$, each platform $i$'s optimal solutions at Nash equilibrium and social optimum can be obtained by solving the following problems, respectively:

$$\max_{\tilde{q}_{i_1},\tilde{q}_{i_2} \geq 0} \tilde{P}_i = (\tilde{q}_{i_1} + \tilde{q}_{i_2})\tilde{F}_i - cN_i = (\tilde{q}_{i_1} + \tilde{q}_{i_2})\left(B(\tilde{Q}) - \beta\left(\tilde{T} + W(\tilde{N}_i^v)\right)\right) - cN_i \quad (45)$$

s.t. Eqs(21),(22),(29)

and

$$\max_{\tilde{q}_{i_1},\tilde{q}_{i_2} \geq 0} \tilde{S} = \int_0^{\tilde{Q}} B(x)dx - \sum_{i=1}^{I} \tilde{q}_{i_1}\beta\left(\tilde{T} + W(\tilde{N}^v)\right) - \sum_{i=1}^{I} \tilde{q}_{i_2}\beta\left(\tilde{T} + W(\tilde{N}_i^v)\right) - cN \quad (46)$$

s.t. Eqs(21),(22),(29)

Due to the complexity of solving the above two optimization problems, we focus on the case in which all passengers opt for the integrator to facilitate our theoretical analysis, i.e., $\tilde{\tau} < \tilde{\tau}_1$. Unless otherwise specified, all passengers in the market with platform-integration are assumed to choose the integrator in the following theoretical derivations. Numerical studies will be conducted to analyze the general mixed market equilibrium with some passengers opting for the integrator and some for individual platforms.

## 4. Evaluating performance of platform-integration

In this section, we compare the optimal solutions between the markets with and without platform-integration at Nash equilibrium as well as social optimum. Our purpose is to evaluate the impacts of the platform-integration, and to quantify the extent to which platform-integration influences realized passenger demand, vehicle utilization rate, platform profit, consumer surplus and social welfare.

### 4.1. Impacts of vehicle fleet size with/without platform-integration at Nash equilibrium/social optimum

Let $U_i$ and $\tilde{U}_i$, denote the vehicle utilization rate of platform $i$ in the markets without and with platform-integration, measured by the fraction of occupied service time, respectively:

$$U_i = \frac{q_i T_i}{N_i}, \quad \forall i \tag{47}$$

and

$$\tilde{U}_i = \frac{(\tilde{q}_{i_1} + \tilde{q}_{i_2})\tilde{T}}{N_i}, \quad \forall i \tag{48}$$

To see the influence of vehicle fleet size on the optimal solutions and system performance metrics, we present the following lemma.

**Lemma 1.** *With Assumptions 1 and 2, assuming $\tilde{\tau} < \tilde{\tau}_1$, it holds:*

(a) *If $N_i > N_j$ and $q_i^{ne}, q_j^{ne} > 0$, then we have $q_i^{ne} > q_j^{ne}$ and $N_i^{v\,ne} > N_j^{v\,ne}$. The same applies to the market with platform-integration.*

(b) *If $N_i > N_j$ and $q_i^{so}, q_j^{so} > 0$, then we have $q_i^{so} > q_j^{so}$ and $N_i^{v\,so} > N_j^{v\,so}$. The same applies to the market with platform-integration.*

(c) *Assuming $N^v \cdot W'(N^v)$ is strictly increasing, if $N_i > N_j$ and $q_i^{so}, q_j^{so} > 0$, then we have $U_i^{so} > U_j^{so}$.*

(d) *In the market with platform-integration, vehicles are utilized at the same degree, i.e., $\tilde{U}_i = \tilde{U}_j, \forall i, j$.*

See **Appendix A** for the proof.

**Lemma 1** states that the platform with a larger fleet size has a greater market share at both Nash equilibrium and social optimum. In addition, the platform with a larger fleet size has a higher vehicle utilization rate at social optimum by assuming $N^v \cdot W'(N^v)$ is a strictly increasing function, for example, $W(N^v) = A(N^v)^{-\kappa}$, which is often used in the literature (Arnott, 1996; Li, et al., 2019).[9] If $\kappa = \frac{1}{2}$, the average waiting and pick-up time of passengers is inversely proportional to the square root of the number of idle vehicles. We should notice that the relation between vehicle fleet size and utilization rate is not monotonic at Nash equilibrium in the market without platform-integration, which will be further discussed in numerical examples.

---

[9] $A$ is an exogenous parameter that encapsulates the factors in the matching technology and depends on the area, vehicle fleet size and vehicle velocity, $\kappa$ is a sensitivity parameter to the number of idle vehicles with $0 < \kappa \leq 1$.

## 4.2. Impacts of platform-integration at Nash equilibrium

We firstly introduce the following lemma, which deals with the total realized demand at Nash equilibrium.

**Lemma 2.** *With Assumptions 1 and 2, assuming $\tilde{\tau} < \bar{\tau}$, then it holds $\tilde{Q}^{ne} > Q^{ne}$ for $I \geq 2$, where $\bar{\tau}$ is given as follows:*

$$\bar{\tau} = B(\tilde{Q}^{ne}) + \frac{1}{I}\tilde{Q}^{ne}B'(\tilde{Q}^{ne}) - \left(B(Q^{ne}) + \frac{1}{I}Q^{ne}B'(Q^{ne})\right) \tag{49}$$

See **Appendix B** for proof.

The assumption that $\tilde{\tau} < \bar{\tau}$, where $\bar{\tau} \leq \tilde{\tau}_1$ measures the difference between the mean marginal total generalized costs perceived by passengers at Nash equilibrium in markets with and without platform-integration. **Lemma 2** implies that platform-integration can increase realized passenger demand at Nash equilibrium if its commission charge is reasonably low; specifically, platform-integration helps to attract more passengers due to reduced generalized costs.

Let $S^{ne}$ and $\tilde{S}^{ne}$ denote the social welfare at Nash equilibrium in the market without and with platform-integration, respectively,

$$S^{ne} = \int_0^{Q^{ne}} B(x)dx - \sum_{i=1}^{I} q_i^{ne}\beta(T_i + W_i^{ne}) - cN \tag{50}$$

$$\tilde{S}^{ne} = \int_0^{\tilde{Q}^{ne}} B(x)dx - \sum_{i=1}^{I} \tilde{q}_i^{ne}\beta(\tilde{T} + \tilde{W}^{ne}) - cN \tag{51}$$

Together with **Lemma 2**, we present the following theorem.

**Theorem 1.** *With Assumptions 1 and 2, assuming $\tilde{\tau} < \bar{\tau}$ and $q_i, \tilde{q}_i > 0$, then at Nash equilibrium, the social welfare in the market with platform-integration is h than that in the market without platform-integration, i.e., $\tilde{S}^{ne} > S^{ne}$ for $I \geq 2$.*

See **Appendix C** for proof.

**Theorem 1** states that platform-integration can improve social welfare at Nash equilibrium. Platform-integration reduces the matching efficiency loss caused by market fragmentation.

However, the impact of platform-integration on the platform profit is uncertain. In the market with platform-integration, there exists two opposite market forces affecting the profit: the positive force by increasing passenger demand and the negative one by reducing trip fare. Whether profit-maximizing platforms benefit from the platform-integration depends on the relative effects of the two opposite forces.

When the supply capacity (i.e., vehicle fleet size) is small or/and there are fewer platforms, the system operates with a very large waiting and pick-up time, then the reduce of waiting and pick-up time due to platform-integration is large due to the convexity of waiting time function, and then attracts more passengers to offset the negative effect of reduced trip fare, leading to a greater profit as well as greater social welfare. In that case, a win-win situation is achieved in which both the passengers and the ride-sourcing platforms are better off due to platform-integration. However, when the supply capacity increases to a certain degree or/and there are more platforms, though with an increase in passenger demand, the big decrease trip fare may lead to an unfavorable result with a shrinking profit for the platforms. In that case, the social welfare is still improved with the platform-integration since the increase in consumer surplus overwhelms the decrease in platform profit. Numerical experiments are conducted in Section 5 for further discussion.

### 4.3. Impact of platform-integration at social optimum

As for the comparison between the total realized demand in the markets with and without platform-integration at social optimum, we present the following lemma.

**Lemma 3.** *With Assumptions 1 and 2, assuming $\tilde{\tau} < \tilde{\tau}_1$ and $\widetilde{N}^{v^{SO}} \geq N_i^{v^{SO}}$, given a waiting time function $W(N^v) = A(N^v)^{-\kappa}$, then it holds $\tilde{Q}^{so} > Q^{so}$ for $I \geq 2$.*

See **Appendix D** for the proof.

The sufficient condition, $\widetilde{N}^{v^{SO}} \geq N_i^{v^{SO}}$, indicates that at social optimum, the total number of idle vehicles in the market with platform-integration is greater than or equal to the number of idle vehicles of each platform in the market without platform-integration. This is a mild assumption that generally holds in actual operations. As aforementioned, the waiting time function $W(N^v) = A(N^v)^{-\kappa}$ is widely used in the literature. From **Lemma 2** and **3**, we generally expect that the total realized demand in the market with platform-integration is greater than that in the market without platform-integration at both Nash equilibrium and social optimum, which implies that the market with platform-integration is more attractive to passengers.

Let $S^{so}$ and $\tilde{S}^{so}$ denote the social welfare at social optimum in the markets without and with platform-integration, respectively,

$$S^{so} = \int_0^{Q^{so}} B(x)dx - \sum_{i=1}^{I} q_i^{so}\beta(T_i + W_i^{so}) - cN \tag{52}$$

and

$$\tilde{S}^{so} = \int_0^{\tilde{Q}^{so}} B(x)dx - \sum_{i=1}^{I} \tilde{q}_i^{so}\beta(\tilde{T} + \widetilde{W}^{so}) - cN \tag{53}$$

Together with **Lemma 3**, we present the following theorem.

**Theorem 2.** *With Assumptions 1 and 2, assuming $\tilde{\tau} < \tilde{\tau}_1$, then at social optimum, the social welfare in the market with platform-integration is greater than or equal to that in the market without platform-integration, i.e., $\tilde{S}^{so} \geq S^{so}$.*

See **Appendix E** for proof.

Together with **Theorem 1**, we find that platform-integration helps to improve social welfare at both Nash equilibrium and social optimum.

## 5. Numerical studies

In this section, we conduct several numerical experiments to illustrate the impacts of platform-integration on the optimal decision variable (i.e., trip fare), key endogenous variables (e.g., realized demand) and system performance metrics (e.g., vehicle utilization rate, platform profit, and social welfare). Specifically, we compare the following two ridesourcing markets with multiple platforms: (1) competitive market without platform-integration; and (2) competitive with platform-integration. In addition, the experiments demonstrate the extent to which the supply (in terms of the number of platforms $I$ and vehicle fleet size $N_i$) affects the performance of platform-integration. Further, we also evaluate the impacts of commission fee charged by the integrator in the general mixed market with some passengers choosing the integrator and others not.

### 5.1. Experimental settings

Consider the following negative exponential demand function:

$$Q = \bar{Q}\exp(-\alpha C) \tag{54}$$

where $\alpha > 0$ is a cost sensitivity parameter, $C$ is the generalized travel cost given by $F + \beta(T + W)$, and $\bar{Q}$ is the potential demand when $C = 0$.

Throughout the numerical studies, we assume the potential demand $\bar{Q} = 1.0 \times 10^5$ (trip/h), cost sensitivity parameter $\alpha = 0.013$ (1/HKD), value of time $\beta = 120$ (HKD/h), unit operating cost per vehicle $c = 50$ (HKD/h) and average trip time $T = 0.4$ (h). The waiting time is assumed to be inversely proportional to the square root of the number of idle vehicles, i.e., $W(N^v) = A/\sqrt{N^v}$, where parameter $A$ is set to be 5 (h). Note that these parameter values are selected with partial reference to previous studies and for illustrative purposes only. In actual operations, one may calibrate the parameters using real data.

**5.2. Impacts of market fragmentation**

In this section, we investigate how market fragmentation (which is reflected by the number of platforms) affects the market performance metrics (e.g. platform profit, social welfare). For illustrative purposes, we assume vehicle fleet sizes are equal across all platforms, i.e., $N_i = N_j, \forall i, j$; the total supply, namely, the sum of the number of vehicles of all platforms, is set to be a constant $2.0 \times 10^4$ (veh), i.e., $N = \sum_i N_i = 2.0 \times 10^4$ (veh); commission fee charged by the integrator is set to be $\tilde{\tau} = 0$ (HKD). By increasing the number of platforms from 1 to 15, we demonstrate how the various market metrics changes with the degree of market fragmentation. The effects of such market fragmentation on the total realized demand, trip fare, total profit and social welfare are shown in **Figure 1**. The green solid line demonstrates how the market metrics change after the implementation of platform-integration, given that the trip fares are unchanged (namely, platforms do not reoptimize their trip fares after platform-integration is introduced). By contrast, the orange solid line indicates the trend of the market metrics with respect to the degree of market fragmentation, under the scenarios where the competing platforms reoptimize their trip fares in response to the implementation of platform-integration such that a new Nash equilibrium is formed.

From **Figure 1**, we can see that the competitive market can be fully efficient regardless of platform-integration as the number of platforms increases to infinity. Namely, the optimal solutions and metrics (e.g., realized demand, profit, and social welfare) at Nash equilibrium of the markets with and without platform-integration converge to those at social optimum of the markets. As aforementioned, the social optimal solutions in the market with platform-integration is independent of the number of platforms (with the market fragmentation cost eliminated), shown as by the constant total demand, profit, and social welfare (see horizontal orange dashed lines in **Figure 1**).

**Total realized demand: Figure 1**-(a) shows that platform-integration helps increase the total realized demand at both Nash equilibrium and social optimum, and the increases become more significant with a large number of platforms. This indicates that, the more of platforms (or the higher the degree of market fragmentation), the more demand brought by platform-integration. Also, it is noted that the total demand in the market without platform-integration at Nash equilibrium initially increase then decrease with the number of platforms. This is because the increase of the number of platforms in the market without platform-integration brings two opposite effects. On one hand, it avoids the platforms from distorting the trip fares from the efficient level (dragging the market from monopoly optimum toward social optimum) by enhancing competitions; on the other hand, it increases the market fragmentation costs. The first effect helps to reduce the trip fare and attracts more passengers, while the second effect helps to increase passengers waiting time and thus discourage passengers from using ride-sourcing services. The implementation of platform-integration is able to maintain the first effect while eliminating the second negative effect; as a result, the total passenger demand of all platforms in the market with platform-integration approaches to the efficient level of passenger demand at social optimum, as the number of platforms increases to infinity. Compared with the green solid line and blue solid line in **Figure 1**-(a), we can see that, even without reoptimizing trip fares, platform-integration can help competing platforms attract more passengers when the number of platforms is small, but not as much as that can be achieved by reoptimizing the trip fares (see orange solid line).

**Trip fare:** From **Figure 1**-(b), we can see that trip fares initially drop greatly with the number of platforms in the markets with and without platform-integration at Nash equilibrium, due to the competition among platforms. As the number of platforms continue increasing, platforms in the market with platform-integration keep reducing trip fares to stimulate a higher demand, while the platforms in the market without platform-integration will even slightly increase trip fares. The latter is because, the matching frictions without platform-integration become larger (i.e., the waiting time becomes larger and the marginal changes of waiting time with respect to a unit increase of supply becomes larger) when the market becomes more fragmented with more platforms. As a result, to offset the increased matching frictions, the for-profit platforms will choose to raise the trip fares to suppress the demand. This counter-intuitive phenomenon is also identified in Zha et al. (2017), who point out that one needs to explore the change of price elasticity of demand and that of matching frictions in order to determine whether competition will lower or raise the price level. By contrast, the platform-integration will eliminate the market fragmentation loss, and thus the platforms will not increase the trip fares to suppress the demand.

**Total profit:** As seen from **Figure 1**-(c), platform-integration may not necessarily improve total profit at Nash equilibrium. Platform-integration has two opposite effects on the total platform profit: (1) it increases total demand by eliminating the market fragmentation cost and reducing the waiting time; (2) it intensifies the competition between platforms and induces them to reduce the trip fares. When there are fewer platforms, the first effect dominates the second effect, namely, platform-integration can greatly increase passenger demand while the platforms still set a comparable trip fare with the situation without platform-integration. As a result, platform-integration can increase platform profit as a whole. By contrast, when there are more platforms, the second effect dominates the first effect, namely, the reduction on trip fare caused by platform-integration is significantly large and can offset the increase of passenger demand. The reason is that, in the markets without platform-integration, the average pick-up time will surge in the situations with many platforms (and thus with severe market fragmentation cost). As aforementioned, at optimal pricing the platforms will set a trip fare to cover drivers' average pick-up and travel time cost plus a monopoly mark-up, therefore, the average trip fare in the market without platform-integration will be higher than that in the market with platform integration (which eliminates the market fragmentation cost). Then we can conclude that platform-integration induce the platforms to set a lower trip fare than would the market without platform-integration, and consequently, reduce the profit of the platforms. Moreover, comparing blue solid line and green solid line in **Figure 1**-(c), we can see that, if the trip fares keep unchanged, profit in the market with platform-integration can be continuously enhanced (as given by the vertical gap) and such enhancement becomes more significant with the number of platforms. However, with reoptimized trip fares under a new Nash equilibrium, profit enhancement becomes less when the number of platforms is larger than 3.

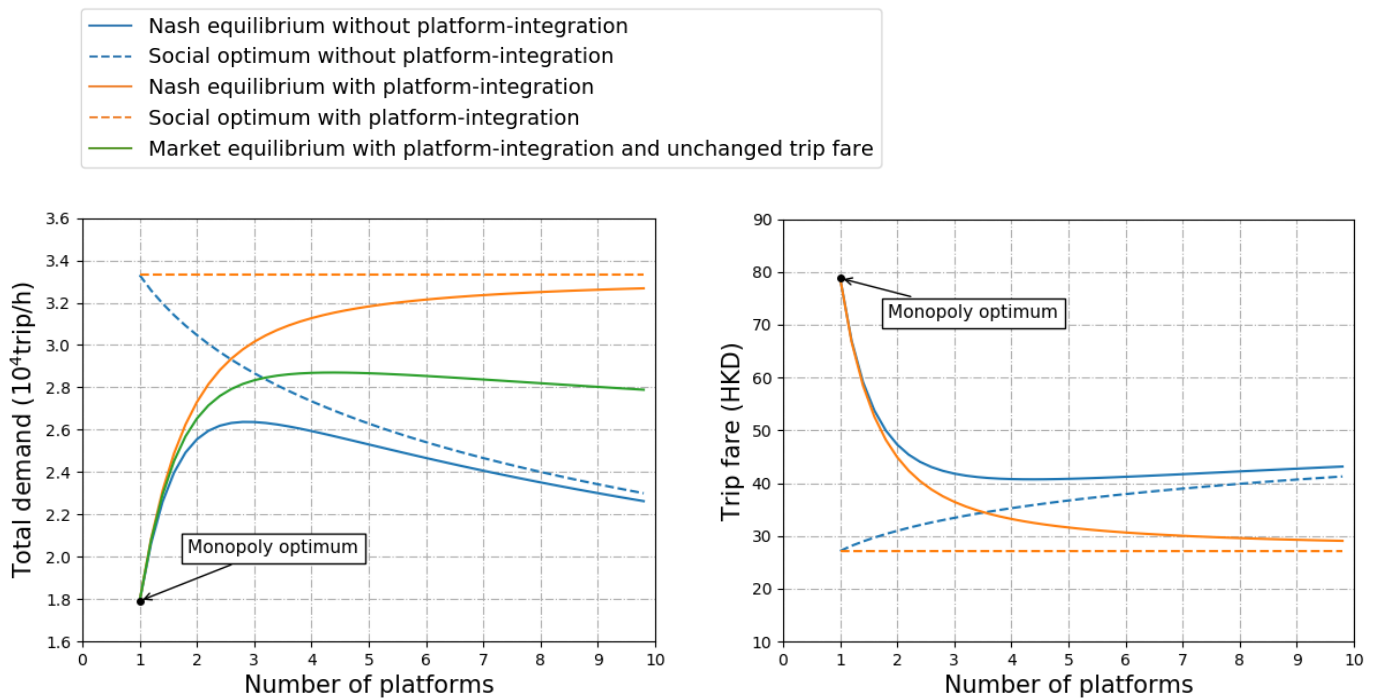

(a). On total realized demand　　　　　　　　　(b). On trip fare

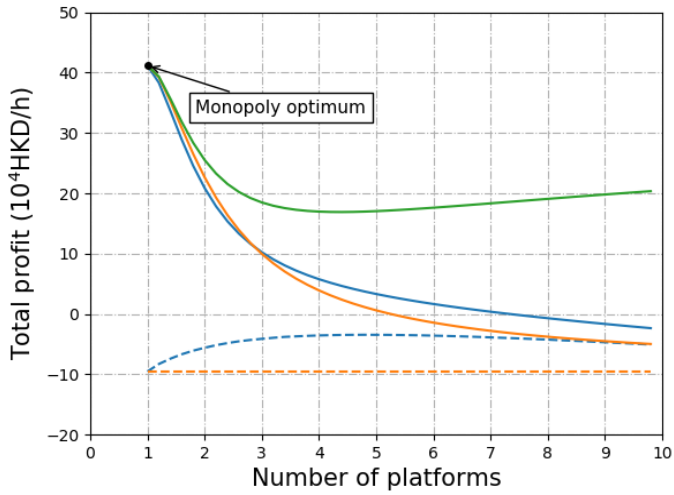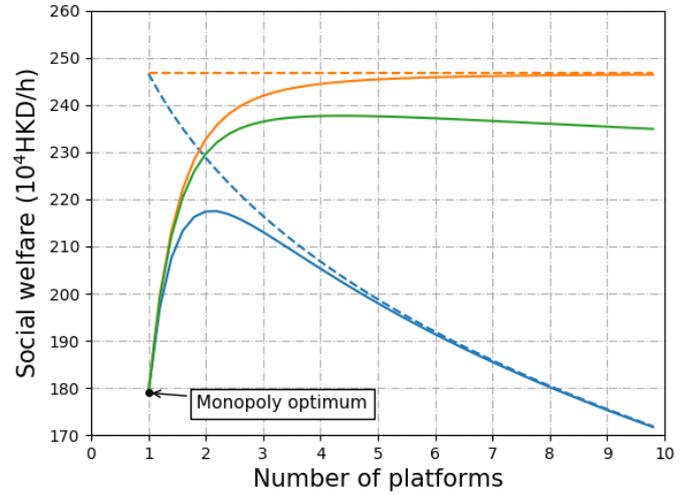

(c). On total profit　　　　　　　　　　　　(d). On social welfare

**Figure 1**. Impacts of the number of platforms on total realized demand, trip fare, total profit and social welfare in the market with/without platform-integration.

**Social welfare:** We can see from **Figure 1**-(d) that platform-integration helps to improve social welfare at both Nash equilibrium and social optimum, which is in accordance with our theoretical results in **Theorem 1** and **2**, and such improvement increases with the number of platforms. This is because, platform-integration can not only maintain competitions and restrain distortion of trip fares, but also reduce the market fragmentation cost caused by competition. By comparing **Figure 1**-(d) and **Figure 1**-(a), we can observe that the curve of change in social welfare is similar to that of total realized demand (which directly dictates the consumer surplus). This is because, when the number of platforms is sufficiently large, the increase in consumer surplus significantly overwhelms the decrease in total profit, giving rise to continuously improved social welfare in the market with platform-integration. In the market without platform-integration, when the number of platforms is sufficiently large, social welfare cuts as a result of the decreases in both total realized demand and total profit. In addition, the market with platform-integration and re-optimization of trip fare can reach more improvement in social welfare than would the market without re-optimization of trip fares.

## 5.3. Impacts of vehicle fleet sizes

This subsection verifies the theoretical findings in Sections 3, 4 and 5, and investigates the impacts of platform-integration on ride-sourcing platforms with different scales (reflected by their vehicle fleet sizes). We assume there are three platforms of different sizes (namely, platform 1, 2 and 3) in the ride-sourcing market, with vehicle fleet size of $N_1 = 500$ (veh), $N_2 = 400$ (veh), and $N_3 = 300$ (veh), respectively; commission fee charged by the integrator is set to be $\tilde{\tau} = 0$ (HKD). We then scale up the vehicle fleet size of each platform by 10% in every instance. The corresponding results of such supply expansion are shown in Figures 2-4, in which the horizontal axis denotes total vehicle fleet size.

**Vehicle utilization rate:** From Figure 2, we can see that vehicle utilization rates of all the three platforms firstly increase and then decrease with fleet sizes in markets with or without platform-integration.[10] Intuitively, at the beginning an increase in vehicle fleet sizes spurs a significant increase in realized passenger demand (see Figure 3-(a)) by reducing passenger waiting time, which is large enough to increase the vehicle utilization; then the realized passenger demand grows slowly with vehicle fleet sizes, which leads to a waste of supply capacity and a lower vehicle utilization. We also find that the platform with the largest fleet size (i.e., platform 1) has the greatest vehicle utilization rate when vehicle fleet size is not sufficiently large, but eventually falls behind in the market without platform-integration at Nash equilibrium (see Figure 2-(a)). By contrast, it always has the greatest vehicle utilization rate in the market without platform-integration at social optimum (see Figure 2-(b)), which is consistent with our theoretical results in **Lemma 1**. The reason is that, in the market without platform-integration, as a result of the increasing return to scale of matching, a platform with larger fleet size has a lower waiting time and attracts more passengers, and the resulting increase in passenger demand is sufficiently large to cause a higher utilization rate, but this marginal effect of increasing supply capacity on passenger demand diminishes more quickly than small platforms. Consequently, the increase in passenger demand fails to keep up with the increase in supply capacity, leading to a lower utilization rate for large platforms. It can also be observed that vehicle utilization rate in the market with platform-integration is larger than utilization rates of platforms of various sizes in the market without platform-integration. This implies that platform-integration can generally help the platforms increase vehicle utilization rate.

---

[10] In the market with platform-integration, given current vehicle fleet sizes and $\tilde{\tau} = 0$, the curves of vehicle utilization rates of all platforms are identical because all passengers opt for the platform-integrator.

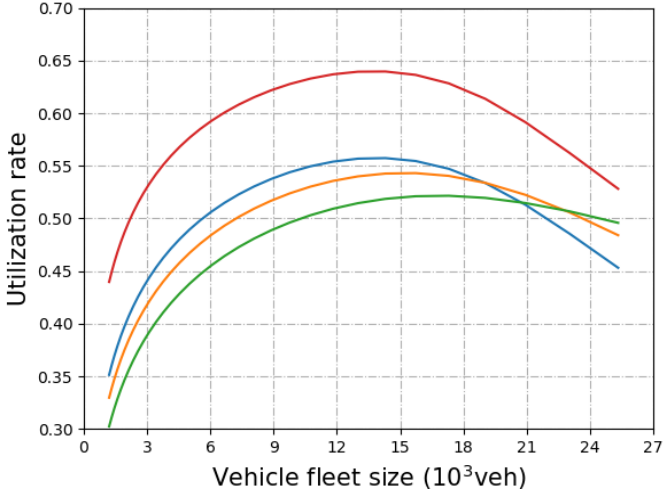
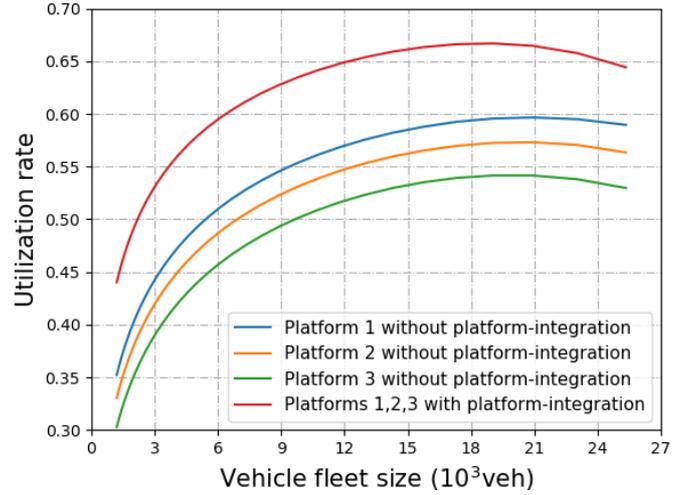

(a). At Nash equilibrium    (b). At social optimum

Figure 2. Impacts of vehicle fleet sizes on vehicle utilization rate in the market with/without platform-integration at Nash equilibrium/social optimum.

**Realized demand:** Next we demonstrate how realized demand changes after the implementation of integration of non-identical platforms. Figure 3-(a) and Figure 3-(b) compare each platform's realized demand between the markets with and without platform-integration, with re-optimized and unchanged trip fares, respectively. We can see from the figure that the realized demand initially increases rapidly with vehicle fleet sizes as a result of increasing return to scale of matching, and then grow slowly because the scale effect is fully exploited. Also, it is noted that platform-integration helps all three platforms of different sizes increase their passenger demands due to reduced passenger waiting time. By comparing the dashed lines in Figure 3-(a) and Figure 3-(b), it can be seen that competitive re-optimization of trip fares after integration attract even more passengers.

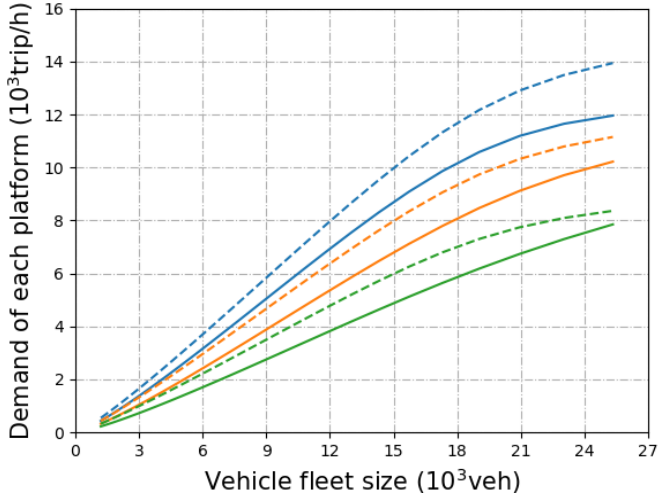 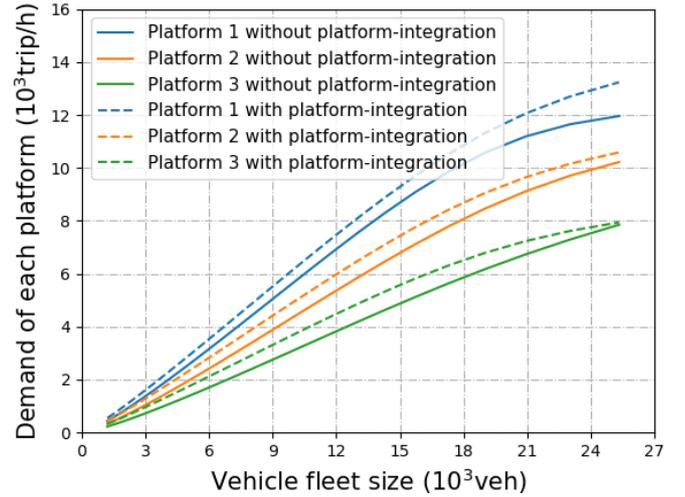

(a) With trip fare re-optimized  (b) With trip fare unchanged

Figure 3. Impacts of vehicle fleet sizes on realized demand in the market with/without platform-integration.

**Profit:** Figure 4-(a) and Figure 4-(b) compare each platform's profit between the markets with and without platform-integration, with re-optimized and unchanged trip fares, respectively. We can see from the figure that the profits initially increase and then decrease with vehicle fleet sizes. When vehicle fleet sizes are small, an increase in supply capacity can significantly boost more demand by reducing passenger waiting time and thus bring a greater profit due to increasing returns to scale of matching function. By comparing the dashed lines (market with integration and re-optimized trip fares) with the solid lines (market without integration) in Figure 4-(a), we can see that platform-integration boosts profits of platforms (especially smaller platforms, e.g., platform 3) when vehicle fleet size is not very large; but it may hurt the profits of platforms (especially larger platforms, e.g., platform 1) with a further increase in fleet sizes, because matching friction is already small and the effect of platform-integration diminishes. As discussed before, whether a platform can gain more profit from platform-integration depends on the extent to which the impetus force from the increased passenger demand mitigates the resistance force from the decreased trip fare. By comparing the dashed lines in Figure 4-(b) (the market with platform-integration and unchanged trip fares) with the dashed lines in Figure 4-(a) (the market with platform-integration and re-optimized trip fares), we find that larger platforms in the market with platform-integration are able to earn more profits in the short run when trip fares are unchanged. However, an iterative decision-process is needed to achieve a stable market equilibrium to evaluate the long-term performance; in this case, the larger platform gains less and even loses profits after the platform-integration.

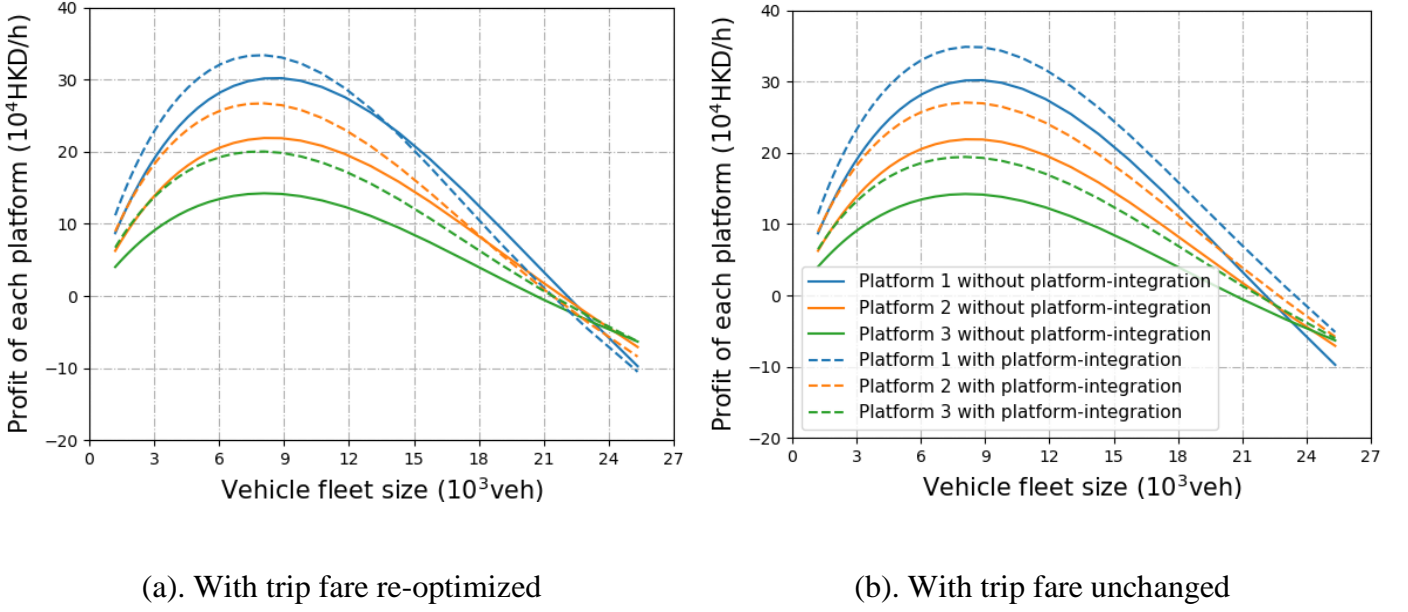

(a). With trip fare re-optimized  (b). With trip fare unchanged

Figure 4. Impacts of vehicle fleet sizes on profit in the markets with/without platform-integration.

**5.4. Impacts of commission fee**

Note that in the numerical studies above, we evaluate the case of $\tilde{\tau} \leq \tilde{\tau}_1$ in which all passengers will choose the integrator in the market with platform-integration. In this section, we present numerical examples to investigate the impacts of commission fee $\tilde{\tau}$ charged by the integrator in the general mixed market equilibrium with some passengers choosing the integrator and others not, i.e., in the case of $\tilde{\tau} \in [\tilde{\tau}_1, \tilde{\tau}_2]$. Assume there are three platforms (namely, platform 1, 2 and 3) in the ride-sourcing market, the trip fare of the three platforms is set to be $F_1 = F_2 = F_3 = 70$ (HKD/trip), and trip fares remain unchanged after platform-integration is implemented, i.e., $\tilde{F}_1 = \tilde{F}_2 = \tilde{F}_3 = 70$ (HKD/trip). Let $(N_1, N_2, N_3)$ denote the supply capacity of the three platforms. We conduct numerical experiments under three different supply, namely, $(2000, 2000, 2000)$, $(3000, 3000, 3000)$, $(3000, 2000, 1000)$ and $(4000, 3000, 2000)$, in units of veh. It is worth nothing that all other parameters are presented in Section 5.1.

Figure 5 shows how total realized demand change with the increase of commission fee in the general mixed market equilibrium, under different supply capacities. The cross at the left (right) end of each curve indicates the critical lower (or upper) threshold of the commission, i.e., $\tilde{\tau}_1$ (or $\tilde{\tau}_2$), below (or beyond) which all (or no) passengers would choose the integrator. In other words, the general mixed market equilibrium can be reached within a range of $\tilde{\tau} \in [\tilde{\tau}_1, \tilde{\tau}_2]$. By comparing orange solid line (red solid line) with blue solid line (green solid line), it is worth noting that the lower and upper bounds of such range generally decrease with vehicle fleet size. This is due to the fact that matching friction is smaller when fleet size is larger given a fixed potential demand, and thus the effect of platform-integration (reduction of market fragmentation) becomes marginal, as a result, the integrator must lower the commission charge in order to retain patronage. In addition, by comparing orange solid line (blue solid) with red solid line (green solid line), we can see that the lower and upper bounds of such range in the case that platforms are of identical size are smaller than the bounds in the case that platforms are of different size, which implies that, when the ridesourcing market has heterogeneous platforms in terms of vehicle capacity, the platform-integration could better attract passengers so that the integrator can afford to increase commission fee to maintain a higher realized demand and achieve a greater profit.

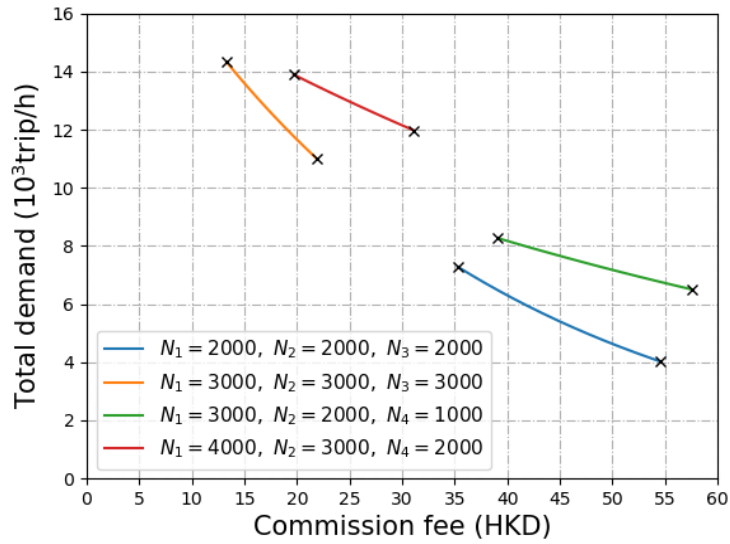

Figure 5. Impacts of commission fee on realized demand given fixed trip fares under different supply capacities.

## 6. Conclusions

In this paper, we investigate a novel business mode in the ride-sourcing market, i.e., the integration of multiple ride-sourcing platforms into one integrator; the integration is referred to as third-party *platform-integration*. We propose mathematical models to describe the ride-sourcing market with multiple competing platforms, and compare vehicle utilization rate, realized demand, profit and social welfare in two markets scenarios: (1) competitive market without platform-integration; and (2) competitive market with platform-integration. The major findings of the study are summarized below.

First, we prove that platform-integration can increase total realized demand and social welfare at both Nash equilibrium and social optimum. Second, we find that platform-integration in general can not only maintain competition and restrain platforms' market power of distorting trip fares, but also reduce the market fragmentation cost caused by competition, leading to the greater social welfare. Third, platform-integration may not increase total profit when the market is too fragmented (i.e., when the number of platforms is large), since the platforms may face cut-throat competition and have to set an extremely low trip fare in the absence of market fragmentation cost (that is eliminated by platform-integration). By contrast, in the market without platform-integration, the presence of the market fragmentation cost will induce the platforms to raise the trip fares to suppress the demand, leading to a relatively high platform profit. Fourth, we find that a win-win situation where both passengers and platforms are better off can be achieved when the supply capacity is not too large. By contrast, platform-integration can still generate greater profits for smaller platforms but may lower profits for larger platforms, when the supply capacity is relatively large. In addition, investigation of platform-integration in general mixed market equilibrium (with some passengers opting for the integrator and others not) is analytically intractable; extensive numerical studies are executed to evaluate its equilibrium. We find that a general mixed market equilibrium can be reached within a certain range of commission fee, and the integrator can afford to set a relatively higher commission fee to maintain a higher realized demand and achieve a greater profit when the total supply level is low or/and platforms are heterogeneous in terms of vehicle capacity.

Our study opens avenues that merit further exploration. To name a few, (1) platform-integration with elastic supply and its impact on the platform profit, consumer surplus, provider (driver) surplus and social welfare; (2) partial platform-integration where passengers have freedom to choose a subset of platforms that offer differentiated ride-sourcing services through the integrator; (3) platform and service mode choice of heterogeneous passengers with different values of time; (4) market fragmentation cost caused by heterogeneous services (e.g., normal ride, luxury ride, and ride-pooling) and the appropriate cross-service integration mechanisms (such as service upgrading) to reduce and even eliminate the cost; (5) mobility-as-a-service (MaaS) with seamless integration of various on-demand services, including ride-sourcing, bike-sharing and on-demand bus services.

**Acknowledgments.** The work described in this paper was supported by a grant from Hong Kong Research Grants Council under project HKUST16208619. This work was also supported by Didi Chuxing. The fourth author gratefully acknowledges the support by the Lee Kong Chian (LKC) Fellowship awarded by Singapore Management University.

**Reference.**

**Web Reference**

**Appendix A. Proof of Lemma 1.**

Define

$$f_i(x) = \frac{N_i - x}{T + W(x)}, x > 0 \qquad (55)$$

where $f_i(x)$ and $x$, respectively, denote $i^{th}$ platform $i$' realized passenger demand $q_i$ and its corresponding number of idle vehicles $N_i^v$ in the market without platform-integration. That is, $f_i(N_i^v) = q_i$ and $dN_i^v/dq_i|_{q_i=q_i^{ne}} = 1/f_i'(N_i^{v\,ne})$.

First-order and second-order derivative with respect $x$

$$f_i'(x) = -\frac{1 + f_i(x)W'(x)}{T + W(x)} \qquad (56)$$

$$f_i''(x) = \frac{2W'(x)(1 + f_i(x)W'(x)) - (N_i - x)W''(x)}{(T + W(x))^2}$$

In the normal region, we have $f_i'(x) < 0$ and $f_i''(x) < 0$. Clearly, if $N_i > N_j$, we have $f_i(x) > f_j(x)$ and $f_i'(x) > f_j'(x)$. Assuming $f_i'(x_i) = f_j'(x_j) = 0$, we obtain $f_i'(x_i) = f_j'(x_j) < f_i'(x_j)$. Based on the fact that $f_i''(x) < 0$, we have $x_i > x_j$, which can be illustrated in Figure .

Define

$$g_i(u) = \frac{1 - u}{T + W(N_i u)} T, u > 0 \qquad (57)$$

where $g_i(u)$ and $u$, respectively, denote the platform $i$'s vehicle utilization rate ($q_i T/N_i$) and ratio of idleness ($N_i^v/N_i$), the latter is defined as the proportion of idle vehicles on the platform.

First-order and second-order derivative with respect $u$

$$g_i'(u) = -\frac{1 + f_i(N_i u)W'(N_i u)}{T + W(N_i u)} T \qquad (58)$$

$$g_i''(u) = \frac{2N_i W'(N_i u)\big(1 + f_i(N_i u)W'(N_i u)\big) - N_i^2(1-u)W''(N_i u)}{\big(T + W(N_i u)\big)^2} T$$

Assuming $u = x/N_i$, $0 < u < 1$, clearly, we have $g_i'(u) = f_i'(x)T$, thus $dN_i^v/dq_i|_{q_i=q_i^{pm}} = T/g_i'(u_i^{pm})$. In the normal region, it holds $g_i'(u) < 0$ and $g_i''(u) < 0$. Define $H_i(u) = 1 + f_i(N_i u)w'(N_i u)$. In the normal region, we have $H_i'(u) = N_i f_i'(N_i u)w'(N_i u) + N_i f_i(N_i u)w''(N_i u) > 0$. Therefore, $H_i(u)$ is a strictly increasing function.

If $N_i > N_j$, we have $x_i > x_j$. Since $xw'(x)$ is a strictly increasing function and $w'(x) < 0$, we further obtain $0 > uN_i w'(uN_i) > uN_j w'(uN_j)$ and $0 < w(uN_i) < w(uN_j)$, leading to $N_i w'(uN_i)\big(T + w(uN_j)\big) > N_j w'(uN_j)\big(T + w(uN_i)\big)$. Then, we can get

$$\frac{N_i(1-u)w'(uN_i)}{\big(T + w(uN_i)\big)} > \frac{N_j(1-u)w'(uN_j)}{\big(T + w(uN_j)\big)}$$

The inequality indicates that if $N_i > N_j$ we have $H_i(u) > H_j(u)$. Assuming $H_i(u_i) = 0$ and $H_j(u_j) = 0$, we have $H_i(u_i) = H_j(u_j) < H_i(u_j)$. Based on the fact that $H_i(u)$ is an increasing function, we have $u_i < u_j$. Since $u_i = x_i/N_i$, from $x_i > x_j$, we can also get $u_i N_i > u_j N_j$.

**Without platform-integration:**

**Lemma 1-(a).** If $q_i^{ne} > 0$ and $q_j^{ne} > 0$, from first-order condition (8), we have

$$B(Q^{ne}) + q_i^{ne} B'(Q^{ne}) = -\beta \frac{dN_i^v}{dq_i}\bigg|_{q_i=q_i^{ne}} \qquad (59)$$

$$B(Q^{ne}) + q_j^{ne} B'(Q^{ne}) = -\beta \frac{dN_j^v}{dq_j}\bigg|_{q_j=q_j^{ne}} \qquad (60)$$

As mentioned, we have $dN_i^v/dq_i < 0$ and $d^2N_i^v/dq_i^2 < 0$ for $\forall i$ in the normal region, then the impact of fleet size can be illustrated in Figure -(a). If $N_i > N_j$, $q_i^{ne} > 0$ and $q_j^{ne} > 0$, let $dN_i^v/dq_i|_{q_i=q_i^{ne*}} = dN_j^v/dq_j|_{q_j=q_j^{ne}}$, then we have $q_i^{ne*} > q_j^{ne}$. However, due to the fact that $B'(Q^{ne}) < 0$, to satisfy Eq (59) and (60), it holds $q_i^{ne*} > q_i^{ne} > q_j^{ne}$. From Figure -(a), we also obtain $N_i^{v\,ne} > N_j^{v\,ne}$. By substituting $U_i^{ne} = q_i^{ne}T/N_i$ and $U_j^{ne} = q_j^{ne}T/N_j$ into Eq (59) and (60), respectively, we get

$$B(Q^{ne}) + \frac{U_i^{ne} N_i}{T} B'(Q^{ne}) = -\beta \frac{dN_i^v}{dq_i}\bigg|_{q_i=q_i^{ne}} \tag{61}$$

$$B(Q^{ne}) + \frac{U_j^{ne} N_j}{T} B'(Q^{ne}) = -\beta \frac{dN_j^v}{dq_j}\bigg|_{q_j=q_j^{ne}} \tag{62}$$

However, the relation between $U_i^{ne}$ and $U_j^{ne}$ is uncertain. In Figure -(b), $u_i^{ne1}$ denotes the case that $U_i^{ne} > U_j^{ne}$, while $u_i^{ne2}$ denotes the case that $U_i^{ne} < U_j^{ne}$.

**Lemma 1-(b).** If $q_i^{so} > 0$ and $q_j^{so} > 0$, from first-order condition (12), we have

$$B(Q^{so}) = -\beta \frac{dN_i^v}{dq_i}\bigg|_{q_i=q_i^{so}}$$

$$B(Q^{so}) = -\beta \frac{dN_j^v}{dq_j}\bigg|_{q_j=q_j^{so}}$$

Similarly, the impact of fleet size at social optimum can be shown in Figure -(c) and (d). That is, if $N_i > N_j$, $q_i^{so} > 0$ and $q_j^{so} > 0$, we have $q_i^{so} > q_j^{so}$, $N_i^{v\,so} > N_i^{v\,so}$ and $U_i^{so} > U_j^{so}$.

**With platform-integration:**

If all passengers opt for the integrator in the market with platform-integration, we have $\tilde{N}_i^v/\tilde{q}_i = \tilde{N}_j^v/\tilde{q}_j$. Using the vehicle conservation constraint (24), we have $(N_i - \tilde{q}_i(\tilde{T} + W(\tilde{N}^v)))/\tilde{q}_i = (N_j - \tilde{q}_j(\tilde{T} + W(\tilde{N}^v)))/\tilde{q}_j$, leading to $\tilde{q}_i\tilde{T}/N_i = \tilde{q}_j\tilde{T}/N_j$, i.e., $\tilde{U}_i = \tilde{U}_j$. From $\tilde{q}_i\tilde{T}/N_i = \tilde{q}_j\tilde{T}/N_j$, we can obtain that passenger demand for each platform from the integrator is proportional to its corresponding vehicle fleet size, i.e., $\tilde{q}_i = \frac{N_i}{N}\tilde{Q}$. Clearly, if $N_i > N_j$, we have $\tilde{q}_i > \tilde{q}_j$.

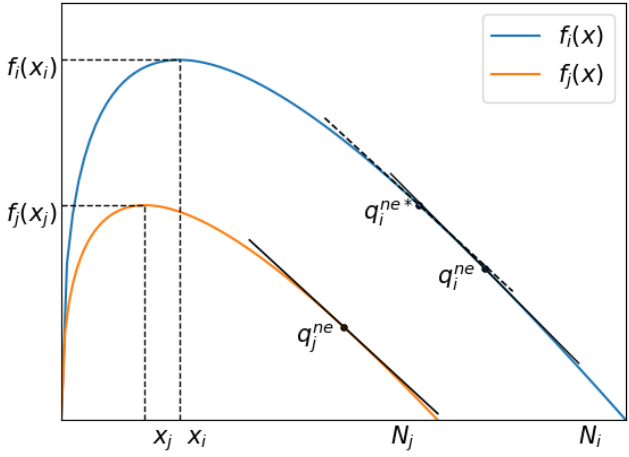
(a). On $q_i$ and $N_i^v$ at Nash equilibrium

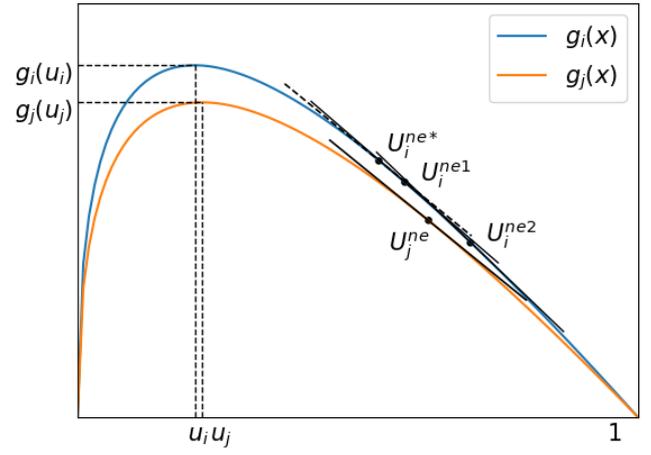
(b). On $U_i$ at Nash equilibrium

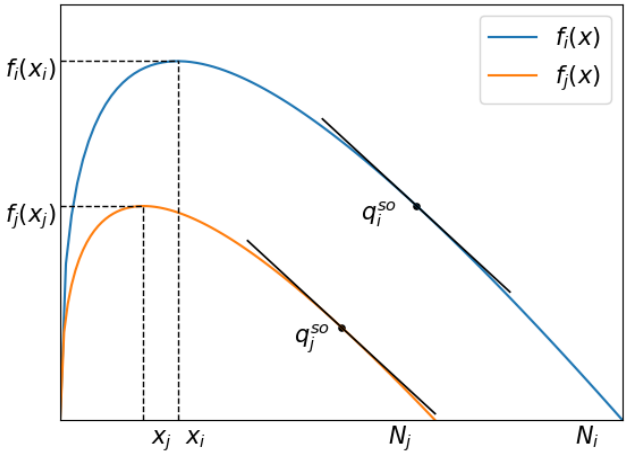
(c). On $q_i$ and $N_i^v$ at social optimum

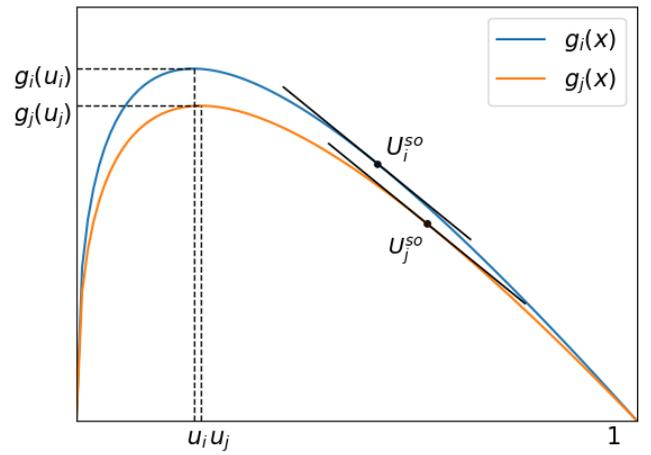
(d). On $U_i$ at social optimum

Figure 6. Impacts of vehicle fleet size at Nash equilibrium/social optimum.

## Appendix B. Proof of Lemma 2.

**Proof.** Clearly, $\tilde{Q}^{ne} = Q^{ne}$ if $I = 1$. Suppose $q_k^{ne} \geq \frac{1}{I} Q^{ne} > 0$, $\tilde{q}_j^{ne} \leq \frac{1}{I} \tilde{Q}^{ne}$, then we have

$$B(Q^{ne}) + \frac{1}{I}Q^{ne}B'(Q^{ne}) \geq B(Q^{ne}) + q_k^{ne}B'(Q^{ne})$$

$$B(\tilde{Q}^{ne}) + \frac{1}{I}\tilde{Q}^{ne}B'(\tilde{Q}^{ne}) \leq B(\tilde{Q}^{ne}) + \tilde{q}_j^{ne}B'(\tilde{Q}^{ne})$$

If $\forall k,j, q_k^{ne} \geq \frac{1}{I}Q^{ne} > 0$, $\tilde{q}_j^{ne} \leq \frac{1}{I}\tilde{Q}^{ne}$, it meets $q_k^{ne} < \tilde{q}_j^{ne}$, clearly we have $Q^{ne} < \tilde{Q}^{ne}$. Otherwise, $\exists k, j$, such that $q_k^{ne} \geq \tilde{q}_j^{ne}$, then we have

$$
\begin{aligned}
B(Q^{ne}) + q_k^{ne}B'(Q^{ne}) &= -\beta \frac{dN_k^v}{dq_k}|_{q_k=q_k^{ne}} \\
&\geq -\beta \frac{dN_k^v}{dq_k}|_{q_k=\tilde{q}_j^{ne}} \\
&> -\frac{N_j}{N}\beta \frac{d\tilde{N}^v}{d\tilde{q}_j}|_{\tilde{q}_j=\tilde{q}_j^{ne}} \\
&\geq B(\tilde{Q}^{ne}) + \tilde{q}_j^{ne}B'(\tilde{Q}^{ne}) - \tilde{\tau}
\end{aligned}
\quad (63)
$$

where from top to bottom, follows from first-order conditions (7) and (8), from $dN_i^v/dq_i < 0$ and $d^2N_i^v/dq_i^2 < 0$ for $\forall i$ in the normal region, from the subsequent proof below, and from first-order condition (32).

Combing the above results yields

$$B(Q^{ne}) + \frac{1}{I}Q^{ne}B'(Q^{ne}) > B(\tilde{Q}^{ne}) + \frac{1}{I}\tilde{Q}^{ne}B'(\tilde{Q}^{ne}) - \tilde{\tau}$$

Assuming $\dot{\tau} = B(\tilde{Q}^{ne}) + \frac{1}{I}\tilde{Q}^{ne}B'(\tilde{Q}^{ne}) - (B(Q^{ne}) + \frac{1}{I}Q^{ne}B'(Q^{ne}))$, then if $\tilde{\tau} < \dot{\tau}$, we have

$$B(Q^{ne}) + \frac{1}{I}Q^{ne}B'(Q^{ne}) > B(\tilde{Q}^{ne}) + \frac{1}{I}\tilde{Q}^{ne}B'(\tilde{Q}^{ne})$$

Based on **Assumption 2-(b)** that $QB(Q)$ is concave, we can derive that $B(Q) + \frac{1}{I}QB'(Q)$ is decreasing. Then we obtain $Q^{ne} < \tilde{Q}^{ne}, I \geq 2$, with the assumption that $\tilde{\tau} < \dot{\tau}$. The proof is completed.

The following part proves the result in (63).

Define

$$\tilde{f}_i(x) = \frac{N_i - \frac{N_i}{N}x}{\tilde{T} + W(x)}, x > 0 \tag{64}$$

First-order and second-order derivative with respect $x$

$$\tilde{f}_i'(x) = -\frac{\frac{N_i}{N} + \tilde{f}_i(x)W'(x)}{\tilde{T} + W(x)} \tag{65}$$

$$\tilde{f}_i''(x) = \frac{2W'(x)\left(1 + \tilde{f}_i(x)W'(x)\right) - \left(N_i - \frac{N_i}{N}x\right)W''(x)}{\left(\tilde{T} + W(x)\right)^2}$$

In the normal region, we have $\tilde{f}_i'(x) < 0$ and $\tilde{f}_i''(x) < 0$. In this study, $x$ and $\tilde{f}_i(x)$ denote total idle vehicles $\tilde{N}^v$ and platform $i$' demand $\tilde{q}_i$ in the market with platform-integration, respectively. Assuming $f_k(x_k) = \tilde{f}_j(x_j) = \tilde{q}_j^{ne}$, then we have $x_j > x_k, I \geq 2$. Based the definitions, comparing $dN_k^v/dq_k|_{q_k=\tilde{q}_j^{ne}}$ and $\frac{N_j}{N}d\tilde{N}^v/d\tilde{q}_j|_{\tilde{q}_j=\tilde{q}_j^{ne}}$ is equivalent to compare $f_k'(x_k)$ and $\frac{N}{N_j}\tilde{f}_j'(x_j)$, that is, if $f_k'(x_k) > \frac{N}{N_j}\tilde{f}_j'(x_j)$, we have $dN_k^v/dq_k|_{q_k=\tilde{q}_j^{ne}} < \frac{N_j}{N}d\tilde{N}^v/d\tilde{q}_j|_{\tilde{q}_j=\tilde{q}_j^{ne}}$.

Using the definition of $\tilde{f}_i'(x)$ and $f_k'(x_k)$, we have

$$\frac{N}{N_j}\tilde{f}_j'(x_j) = -\frac{1 + \frac{N}{N_j}\tilde{f}_j(x_j)W'(x_j)}{\tilde{T} + W(x_j)}$$

$$> -\frac{1 + \tilde{f}_j(x_j)W'(x_j)}{\tilde{T} + W(x_j)}$$

$$> -\frac{1 + \tilde{f}_j(x_j)W'(x_k)}{\tilde{T} + W(x_k)}$$

$$= -\frac{1 + f_k(x_k)W'(x_k)}{\tilde{T} + W(x_k)}$$

$$= f_k'(x_k)$$

where from top to bottom, follows from the definition (56) of $\tilde{f}_j'(x_j)$, from the fact that $N/N_j \geq 1$, from the fact that $x_k > x_j$, from the fact that $f_k(x_k) = \tilde{f}_j(x_j)$, and from the definition (65) of $f_k'(x_k)$.

Thus, we have $\frac{N}{N_j}\tilde{f}_j'(x_j) < f_k'(x_k)$, which leads to $dN_k^v/dq_k|_{q_k=\tilde{q}_j^{ne}} < \frac{N_j}{N}d\widetilde{N}^v/d\tilde{q}_j|_{\tilde{q}_j=\tilde{q}_j^{ne}}$, and we prove Eq(63).

**Appendix C. Proof of Theorem 1.**

Based on the social welfare at Nash equilibrium in the market with platform-integration given in Eq(51) and without platform-integration given in Eq(50), we obtain

$$\tilde{S}^{ne} - S^{ne} = \int_{Q^{ne}}^{\tilde{Q}^{ne}} B(x)dx - (\sum_{i=1}^{I}\tilde{q}_i^{ne}\beta(\tilde{T} + \widetilde{W}^{ne}) - \sum_{i=1}^{I}q_i^{ne}\beta(T + W_i^{ne}))$$

Substituting vehicle conservation constrains (2) and (24) into the above equation yields

$$\tilde{S}^{ne} - S^{ne} = \int_{Q^{ne}}^{\tilde{Q}^{ne}} B(x)dx + \sum_{i=1}^{I}\beta(\frac{N_i}{N}\widetilde{N}^{v^{ne}} - N_i^{v^{ne}})$$

Because $B(x)$ is convex and strictly decreasing, we have

$$\int_{Q^{ne}}^{\tilde{Q}^{ne}} B(x)dx \geq (\tilde{Q}^{ne} - Q^{ne})B(\tilde{Q}^{ne})$$

Based on the definition of $\tilde{f}_i(x)$ (64) and $f_i(x)$ (55), we have $\tilde{q}_i^{ne} = \tilde{f}_i(\widetilde{N}^{v^{ne}})$ and $q_i^{ne} = f_i(N_i^{v^{ne}})$. Suppose $q_i'^{ne} = \tilde{f}_i\left(\frac{N}{N_i}N_i^{v^{ne}}\right)$, then it holds $q_i'^{ne} \geq q_i^{ne}$.

Because we have $d\widetilde{N}^v/d\tilde{q}_i < 0$ and $d^2\widetilde{N}^v/d\tilde{q}_i^2 < 0$ in the normal region as well as the fact that $q_i'^{ne} \geq q_i^{ne}$, then we have

$$\sum_{i=1}^{I}\beta(\frac{N_i}{N}\widetilde{N}^{v^{ne}} - N_i^{v^{ne}}) \geq \sum_{i=1}^{I}\frac{N_i}{N}\beta(\tilde{q}_i^{ne} - q_i'^{ne})\frac{d\widetilde{N}^v}{d\tilde{q}_i}|_{\tilde{q}_i=\tilde{q}_i^{ne}} \geq \sum_{i=1}^{I}\frac{N_i}{N}\beta(\tilde{q}_i^{ne} - q_i^{ne})\frac{d\widetilde{N}^v}{d\tilde{q}_i}|_{\tilde{q}_i=\tilde{q}_i^{ne}}$$

Combining the results yields

$$\tilde{S}^{ne} - S^{ne} \geq \sum_{i=1}^{I}(\tilde{q}_i^{ne} - q_i^{ne})(B(\tilde{Q}^{ne}) + \frac{N_i}{N}\beta\frac{d\widetilde{N}^v}{d\tilde{q}_i}|_{\tilde{q}_i=\tilde{q}_i^{ne}})$$

Substituting first-order conditions (31) and (32) into the above inequation obtains

$$\tilde{S}^{ne} - S^{ne} \geq -B'(\tilde{Q}^{ne}) \sum_{i=1}^{I} (\tilde{q}_i^{ne} - q_i^{ne}) \tilde{q}_i^{ne} + \tilde{\tau}(\tilde{Q}^{ne} - Q^{ne}) \qquad (66)$$

The minimum value of $\sum_{i=1}^{I}(\tilde{q}_i^{ne} - q_i^{ne})\tilde{q}_i^{ne}$ is $(\tilde{Q}^{ne} - Q^{ne})\tilde{Q}^{ne}/I$, which follows readily from the first-order conditions of the following problem.

$$\min Z = \sum_{i=1}^{I} (\tilde{q}_i^{ne} - q_i^{ne}) \tilde{q}_i^{ne}$$

subject to

$$Q^{ne} = \sum_{i=1}^{I} q_i^{ne}, \quad q_i^{ne} > 0$$

$$\tilde{Q}^{ne} = \sum_{i=1}^{I} \tilde{q}_i^{ne}, \quad \tilde{q}_i^{ne} > 0$$

Then we obtain

$$\tilde{S}^{ne} - S^{ne} \geq -\frac{B'(\tilde{Q}^{ne})(\tilde{Q}^{ne} - Q^{ne})\tilde{Q}^{ne}}{I} + \tilde{\tau}(\tilde{Q}^{ne} - Q^{ne})$$

Based on the fact that $\tilde{Q}^{ne} > Q^{ne}$ from **Lemma 2** and the fact that $B'(\tilde{Q}^{ne}) < 0$, we arrive at $\tilde{S}^{ne} > S^{ne}, I \geq 2$.

With regard to the profit, define

$$P^{ne} = \sum_{i=1}^{I} q_i^{ne}(B(Q^{ne}) - \beta(T + W_i^{ne}) - \tilde{\tau}) - cN$$

$$\tilde{P}^{ne} = \sum_{i=1}^{I} \tilde{q}_i^{ne}(B(\tilde{Q}^{ne}) - \beta(\tilde{T} + \tilde{W}^{ne}) - \tilde{\tau}) - cN$$

Then we have

$$\tilde{P}^{ne} - P^{ne} = \tilde{Q}^{ne} B(\tilde{Q}^{ne}) - Q^{ne} B(Q^{ne}) + \sum_{i=1}^{I} \beta \left( \frac{N_i}{N} \tilde{N}^{vne} - N_i^{vne} \right) - \tilde{\tau}(\tilde{Q}^{ne} - Q^{ne})$$

$$= (\tilde{Q}^{ne} - Q^{ne}) B(\tilde{Q}^{ne}) + \sum_{i=1}^{I} \beta \left( \frac{N_i}{N} \tilde{N}^{vne} - N_i^{vne} \right) + Q^{ne} \left( B(\tilde{Q}^{ne}) - B(Q^{ne}) \right) - \tilde{\tau}(\tilde{Q}^{ne} - Q^{ne})$$

$$\geq -B'(\tilde{Q}^{ne}) \sum_{i=1}^{I} (\tilde{q}_i^{ne} - q_i^{ne})\tilde{q}_i^{ne} + Q^{ne}\left(B(\tilde{Q}^{ne}) - B(Q^{ne})\right)$$

**Appendix D. Proof of Lemma 3.**

Clearly, we have $Q^{so} = \tilde{Q}^{so}, I = 1$.

consider the following waiting time function

$$W(N^v) = A(N^v)^{-\kappa} \tag{67}$$

First-order derivative with respect $N_v$

$$W'(N^v) = -\kappa A(N^v)^{-(\kappa+1)} \tag{68}$$

Define

$$\tilde{f}_I(x) = \frac{N - x}{\tilde{T} + W(x)}, x > 0 \tag{69}$$

First-order and second-order derivative with respect $x$

$$\tilde{f}_I'(x) = -\frac{1 + \tilde{f}_I(x)W'(x)}{\tilde{T} + W(x)} \tag{70}$$

$$\tilde{f}_I''(x) = \frac{2W'(x)\left(1 + \tilde{f}_I(x)W'(x)\right) - (N - x)W''(x)}{\left(\tilde{T} + W(x)\right)^2}$$

Suppose $q_j^{so} \geq \frac{1}{I}Q^{so} > 0$, based on the definitions (55) and (69), we have $Q^{so} = \tilde{f}_I(\tilde{N}^{v\,so})$ and $q_j^{so} = f_j(N_j^{v\,so})$. If $\forall j, q_j^{so} \geq \frac{1}{I}Q^{so}$, it meets $\tilde{Q}^{so} > Iq_j^{so}$, then clearly we have $\tilde{Q}^{so} > Q^{so}$.

Otherwise, if $\exists j, \tilde{Q}^{so} \leq Iq_j^{so}$, then it holds

$$1 + \tilde{f}_I(\tilde{N}^{v\,so})W'(\tilde{N}^{v\,so}) \geq 1 + If_j(N_j^{v\,so})W'(\tilde{N}_j^{v\,so})$$

$$= 1 - \kappa A I f_j(N_j^{v\,so})(\tilde{N}_j^{v\,so})^{-(\kappa+1)}$$

$$> 1 - \kappa A f_j(N_j^{v\,so})(N_j^{v\,so})^{-(\kappa+1)}$$

$$= 1 + f_j(N_j^{v\,so}) W'(N_j^{v\,so})$$

where from top to bottom, follows from the fact that $q_j^{so} \geq \frac{1}{I} Q^{so}$ and $\widetilde{N}^{v\,so} \geq \widetilde{N}_j^{v\,so}$, from the definition (70) of $W'(\widetilde{N}_j^{v\,so})$, from the assumption that $\widetilde{N}^{v\,so} \geq N_j^{v\,so}$ and the fact that $I \geq 2$, and from the definition (70) of $W'(N_j^{v\,so})$.

Using first-order conditions (12) and (44), the definitions (56) and (70), we have

$$B(\tilde{Q}^{so}) = -\beta \frac{d\widetilde{N}^v}{d\tilde{Q}}|_{\tilde{Q}=\tilde{Q}^{so}} = -\beta \frac{1}{\tilde{f}_I'(\widetilde{N}^{v\,so})} = \frac{\beta \left(T + W(\widetilde{N}^{v\,so})\right)}{1 + \tilde{f}_I(\widetilde{N}^{v\,so}) W'(\widetilde{N}^{v\,so})}$$

$$B(Q^{so}) = -\beta \frac{dN_i^v}{dq_j}|_{q_j=q_j^{so}} = -\beta \frac{1}{f_j'(N_j^{v\,so})} = \frac{\beta \left(T + W(N_j^{v\,so})\right)}{1 + f_j(N_j^{v\,so}) W'(N_j^{v\,so})}$$

Finally, we arrive at $B(\tilde{Q}^{so}) < B(Q^{so})$ based on the assumption that $\widetilde{N}^{v\,so} \geq N_j^{v\,so}$. Because $B(\cdot)$ is strictly decreasing, we obtain $\tilde{Q}^{so} > Q^{so}$.

**Appendix E. Proof of Theorem 2.**

At social optimum, proceeding as the same as the proof of **Theorem 1** in **Appendix C**, we can prove

$$\tilde{S}^{so} - S^{so} \geq \sum_{i=1}^{I} (\tilde{q}_i^{so} - q_i^{so})\left(B(\tilde{Q}^{so}) + \frac{N_i}{N} \beta \frac{d\widetilde{N}^v}{d\tilde{q}_i}|_{\tilde{q}_i=\tilde{q}_i^{so}}\right)$$

Based on first-order conditions (35) and (36) and the fact $\tilde{q}_i^{so} > 0, \forall i$, we have

$$B(\tilde{Q}^{so}) + \frac{N_i}{N} \beta \frac{d\widetilde{N}^v}{d\tilde{q}_i}|_{\tilde{q}_i=\tilde{q}_i^{so}} = 0$$

Thus, we arrive at $\tilde{S}^{so} \geq S^{so}$.